\documentclass[12pt]{article}
 \usepackage{epsfig}
\newcommand{\be}{\begin{equation}}
\newcommand{\ee}{\end{equation}}
\newcommand{\beq}{\begin{equation}}
\newcommand{\eeq}{\end{equation}}
\newcommand{\bea}{\begin{eqnarray}}
\newcommand{\eea}{\end{eqnarray}}
\newcommand{\nn}{\nonumber}

\begin{document}
\baselineskip=15.5pt
\pagestyle{plain}
\setcounter{page}{1}


\def\del{{\partial}}
\def\vev#1{\left\langle #1 \right\rangle}
\def\cn{{\cal N}}
\def\co{{\cal O}}
\newfont{\Bbb}{msbm10 scaled 1200}     
\newcommand{\mathbb}[1]{\mbox{\Bbb #1}}
\def\IC{{\mathbb C}}
\def\IR{{\mathbb R}}
\def\IZ{{\mathbb Z}}
\def\RP{{\bf RP}}
\def\CP{{\bf CP}}
\def\Poincare{{Poincar\'e }}
\def\tr{{\rm tr}}
\def\tp{{\tilde \Phi}}

\def\TL{\hfil$\displaystyle{##}$}
\def\TR{$\displaystyle{{}##}$\hfil}
\def\TC{\hfil$\displaystyle{##}$\hfil}
\def\TT{\hbox{##}}
\def\HLINE{\noalign{\vskip1\jot}\hline\noalign{\vskip1\jot}}
\def\seqalign#1#2{\vcenter{\openup1\jot
  \halign{\strut #1\cr #2 \cr}}}
\def\lbldef#1#2{\expandafter\gdef\csname #1\endcsname {#2}}
\def\eqn#1#2{\lbldef{#1}{(\ref{#1})}%
\begin{equation} #2 \label{#1} \end{equation}}
\def\eqalign#1{\vcenter{\openup1\jot
    \halign{\strut\span\TL & \span\TR\cr #1 \cr
   }}}
\def\eno#1{(\ref{#1})}
\def\href#1#2{#2}
\def\half{{1 \over 2}}

\def\ads{{\it AdS}}
\def\adsp{{\it AdS}$_{p+2}$}
\def\cft{{\it CFT}}

\newcommand{\ber}{\begin{eqnarray}}
\newcommand{\eer}{\end{eqnarray}}

\newcommand{\beqar}{\begin{eqnarray}}
\newcommand{\cN}{{\cal N}}
\newcommand{\cO}{{\cal O}}
\newcommand{\cA}{{\cal A}}
\newcommand{\cT}{{\cal T}}
\newcommand{\cF}{{\cal F}}
\newcommand{\cC}{{\cal C}}
\newcommand{\cR}{{\cal R}}
\newcommand{\cW}{{\cal W}}
\newcommand{\eeqar}{\end{eqnarray}}
\newcommand{\tht}{\thteta}
\newcommand{\lm}{\lambda}\newcommand{\Lm}{\Lambda}
\newcommand{\eps}{\epsilon}


\newcommand{\nonu}{\nonumber}
\newcommand{\oh}{\displaystyle{\frac{1}{2}}}
\newcommand{\dsl}
  {\kern.06em\hbox{\raise.15ex\hbox{$/$}\kern-.56em\hbox{$\partial$}}}
\newcommand{\id}{i\!\!\not\!\partial}
\newcommand{\as}{\not\!\! A}
\newcommand{\ps}{\not\! p}
\newcommand{\ks}{\not\! k}
\newcommand{\D}{{\cal{D}}}
\newcommand{\dv}{d^2x}
\newcommand{\Z}{{\cal Z}}
\newcommand{\N}{{\cal N}}
\newcommand{\Dsl}{\not\!\! D}
\newcommand{\Bsl}{\not\!\! B}
\newcommand{\Psl}{\not\!\! P}
\newcommand{\eeqarr}{\end{eqnarray}}
\newcommand{\ZZ}{{\rm \kern 0.275em Z \kern -0.92em Z}\;}

                                                                                                    
\def\del{{\delta^{\hbox{\sevenrm B}}}} \def\ex{{\hbox{\rm e}}}
\def\azb{A_{\bar z}} \def\az{A_z} \def\bzb{B_{\bar z}} \def\bz{B_z}
\def\czb{C_{\bar z}} \def\cz{C_z} \def\dzb{D_{\bar z}} \def\dz{D_z}
\def\im{{\hbox{\rm Im}}} \def\mod{{\hbox{\rm mod}}} \def\tr{{\hbox{\rm Tr}}}
\def\ch{{\hbox{\rm ch}}} \def\imp{{\hbox{\sevenrm Im}}}
\def\trp{{\hbox{\sevenrm Tr}}} \def\vol{{\hbox{\rm Vol}}}
\def\rl{\Lambda_{\hbox{\sevenrm R}}} \def\wl{\Lambda_{\hbox{\sevenrm W}}}
\def\fc{{\cal F}_{k+\cox}} \def\vev{vacuum expectation value}
\def\nodiv{\mid{\hbox{\hskip-7.8pt/}}}
\def\ie{{\em i.e.}}
\def\ie{\hbox{\it i.e.}}

\def\CC{{\mathchoice
{\rm C\mkern-8mu\vrule height1.45ex depth-.05ex
width.05em\mkern9mu\kern-.05em}
{\rm C\mkern-8mu\vrule height1.45ex depth-.05ex
width.05em\mkern9mu\kern-.05em}
{\rm C\mkern-8mu\vrule height1ex depth-.07ex
width.035em\mkern9mu\kern-.035em}
{\rm C\mkern-8mu\vrule height.65ex depth-.1ex
width.025em\mkern8mu\kern-.025em}}}
                                                                                                    
\def\RR{{\rm I\kern-1.6pt {\rm R}}}
\def\NN{{\rm I\!N}}
\def\ZZ{{\rm Z}\kern-3.8pt {\rm Z} \kern2pt}
\def\IB{\relax{\rm I\kern-.18em B}}
\def\ID{\relax{\rm I\kern-.18em D}}
\def\II{\relax{\rm I\kern-.18em I}}
\def\IP{\relax{\rm I\kern-.18em P}}
\newcommand{\CS}{{\scriptstyle {\rm CS}}}
\newcommand{\CSs}{{\scriptscriptstyle {\rm CS}}}
\newcommand{\rc}{\nonumber\\}
\newcommand{\bear}{\begin{eqnarray}}
\newcommand{\eear}{\end{eqnarray}}
\newcommand{\W}{{\cal W}}
\newcommand{\F}{{\cal F}}
\newcommand{\x}{{\cal O}}
\newcommand{\LL}{{\cal L}}
                                                                                                    
\def\mani{{\cal M}}
\def\calo{{\cal O}}
\def\calb{{\cal B}}
\def\calw{{\cal W}}
\def\calz{{\cal Z}}
\def\cald{{\cal D}}
\def\calc{{\cal C}}
\def\to{\rightarrow}
\def\ele{{\hbox{\sevenrm L}}}
\def\ere{{\hbox{\sevenrm R}}}
\def\zb{{\bar z}}
\def\wb{{\bar w}}
\def\nodiv{\mid{\hbox{\hskip-7.8pt/}}}
\def\menos{\hbox{\hskip-2.9pt}}
\def\dr{\dot R_}
\def\drr{\dot r_}
\def\ds{\dot s_}
\def\da{\dot A_}
\def\dga{\dot \gamma_}
\def\ga{\gamma_}
\def\dal{\dot\alpha_}
\def\al{\alpha_}
\def\cl{{closed}}
\def\cls{{closing}}
\def\vev{vacuum expectation value}
\def\tr{{\rm Tr}}
\def\to{\rightarrow}
\def\too{\longrightarrow}

\def\S{\Sigma}
\def\a{\alpha}
\def\b{\beta}
\def\c{\gamma}
\def\d{\delta}
\def\e{\epsilon}           
\def\f{\phi}               
\def\vf{\varphi}  \def\tvf{\tilde{\varphi}}
\def\g{\gamma}
\def\h{\eta}
\def\i{\iota}
\def\j{\psi}
\def\k{\kappa}                    
\def\l{\lambda}
\def\m{\mu}
\def\n{\nu}
\def\o{\omega}  \def\w{\omega}
\def\q{\theta}  \def\th{\theta}                  
\def\r{\rho}                                     
\def\s{\sigma}                                   
\def\t{\tau}
\def\u{\upsilon}
\def\x{\xi}
\def\z{\zeta}
\def\pt{\tilde{\varphi}}
\def\tt{\tilde{\theta}}
\def\la{\label}                                                                                                    
\def\6{\partial}
\def\wg{\wedge}

%
                                                                                                    
\newfont{\namefont}{cmr10}
\newfont{\addfont}{cmti7 scaled 1440}
\newfont{\boldmathfont}{cmbx10}
\newfont{\headfontb}{cmbx10 scaled 1728}
\renewcommand{\theequation}{{\rm\thesection.\arabic{equation}}}
\begin{titlepage}

\begin{center} \Large \bf Glueballs of Super Yang-Mills from wrapped 
branes 

\end{center}

\vskip 0.3truein
\begin{center}
Elena C\'aceres$^{\dagger}$\footnote{elenac@zippy.ph.utexas.edu} and 
Carlos N\'u\~nez$^{*}$ 
\footnote{nunez@lns.mit.edu}
\vspace{0.3in}

${}^\dagger$ CINVESTAV \\Apdo. 14-740, 07000 Mexico D.F, Mexico. \\
{\small and}\\
Theory Group, Department of Physics, University of Texas at Austin\\
Austin, TX 78712, USA\\
 \vspace{0.3in}
${}^{*}$ Center for Theoretical Physics, Massachusetts Institute of
Technology \\
Cambridge, MA 02139, USA\\

\vspace{0.3in}

\end{center}
\vskip 1truein

{\bf ABSTRACT:}
In this paper we study qualitative features  of glueballs in $\cN=1$ SYM
for models of  wrapped branes in  IIA and IIB backgrounds. The scalar mode, $0^{++}$ 
is found to be a mixture of the dilaton and the internal part of the metric. 
We carry out the numerical study of the IIB background. 
The potential found exhibits a mass gap and produces a discrete spectrum 
without any cut-off. We propose a 
regularization procedure needed to make these states normalizable.

\vskip2.6truecm
\vspace{0.3in}
\leftline{UTTG-05-05 }
\vspace{0.2in}
\leftline{MIT-CPT 3651}
\vspace{0.2in}
\leftline{hep-th/0506051 }
\smallskip
\end{titlepage}
\setcounter{footnote}{0}

\section{Introduction and General Ideas}
During  the sixties the study of hadronic 
particles was a mainstream area of theoretical Physics.
The Regge trajectories were proposed and  
by the systematic study of scattering of Hadrons, the dualities 
between the $s$ and $t$ channels amplitudes were investigated and 
nicely realized by the Veneziano amplitude, 

The dual models, precursors of the modern String Theory, were one of 
the best candidates to explain the relevant Physics and provided the 
tools to 
explore the issues mentioned above. 
Despite some success, the experiments realized during the late 
sixties involving 
scatterings of large $s$ and large $t$ Mandelstam variables, 
but keeping fixed $\frac{s}{t}$ (fixed angle processes), gave 
results that showed that the amplitude was falling as a power law 
instead of the exponential law predicted by the dual models.
This situation, together with the advent of QCD (that correctly predicted 
the scaling for fixed angle scattering and many other  
things), lead to the 
demise of the dual models for the study of hadronic 
Physics. 

Even when the original motivation to study dual models 
momentarily dissapeared, their rich structure kept 
many physicists interested and, after some technical subtleties were 
understood, this was the beginning of String Theory.

As is  known, after thirty years, string theorists have come back to 
the study of 
problems related to the hadronic world. Indeed, guided by the 
Maldacena Conjecture \cite{Maldacena:1997re} and their refinements 
\cite{Gubser:1998bc}, there have been very many interesting 
achievements in the area. They correspond to field theories 
with different amount of SUSY (including no SUSY) but ``very 
similar'' to QCD. The important 
point is that many characteristic features of QCD, like confinement, 
chiral symmetry breaking, etc; have been understood based on dual 
String theory backgrounds.

In this paper, we study glueballs in some of the models mentioned 
above as being ``very similar'' to QCD (even when the models 
we will deal with here and those available in the literature, 
perhaps are not in the same universality class 
of QCD). Let us motivate a little bit the study of these  
glue-composed excitations.

We know that the main distinction between a field theory in a 
confining phase and the same field theory in the Higgs phase is the 
presence of Regge trajectories, that do not occur in theories with 
Coulomb of Yukawa interactions. These Regge  trajectories appear 
when plotting the spin $J$ and the squarred mass $m^2$ of the 
excitation, thus giving 
relation of the form $J=\alpha' m^2 + \alpha_0,$ with 
$\alpha'\approx (1 GeV)^{-2}$; this relation does not have in 
principle, an upper bound in $J,m$. 
It is due to these infinite number of  Regge resonances, 
being interchanged in the $s,t$ channels of any hadron scattering 
that the beautiful structure of duality appeared in the models above 
mentioned. The glueballs should be some of these Regge excitations (making 
up a full trajectory if mixing with quarks is neglected) 
and this is a possible motivation to study them.

From a modern QCD perspective,
it is known that the cloud of gluons is what logically connects 
between a current quark (with mass of a few $MeV$) to a constituent 
quark, with mass of around 300 $MeV$. Since glue is part of the 
hadronic matter, we can consider color singlet composites of the 
form $q\bar{q} g, \; gg, \; ggg$ (apart from the mesons, 
baryons and exotics). The glueballs are composites made out of 
constituent glue, with no quark content.
Of course, since we live in a world with quarks,  one might think 
that the proposal of pure glue objects is impossible to study, 
because the quarks should run in loops when doing corrections to 
the operators, that leads to glueballs mixing with mesons, 
rendering the object not-pure glue. But lattice theorist 
(working in the quenched approximation) are not stopped by this. Indeed they 
took advantage of the limitation  and  have taught us many things 
about glueballs. 

Among the things that Lattice showed about QCD glueballs, we can 
mention the facts that:
\begin{itemize}
\item{there is a bound state spectrum} 
\item{the 
lightest glueball is a scalar} \item{ the next is a tensor, 1.6 
times 
heavier} \item{ the mass of the lightest glueball should be around 
1630 $MeV$;}
\end{itemize} 
see for example \cite{Teper:1998kw}
for a nice and clear review of these results.

How are these lattice predictions experimentally checked?
Experimentalist look for processes rich in glue production, like the 
$J/\psi$ decay, where the $c, \bar{c}$ quarks annihilate into gluons. 
Other process might be the $p\bar{p}$ annihilation, in this case the 
idea is that the quarks and anti-quarks in the initial hadrons 
annihilate completely, producing glue that later decays into hadrons.  
There are many glueballs candidates. One of them seems to be well 
established and is called $f_0(1500)$ with a width of $112\; MeV$ 
\cite{Amsler:1995gf}.

From the string theory view point, using the Maldacena duality for 
the case of confining backgrounds, the study of glueballs (in the 
field theory dual to the background) proceeds by finding bound 
states for the fluctuations of the supergravity fields. Basically, 
the idea is to fluctuate all the fields in a given IIA, or IIB 
solution dual to a confining field theory, and  linearizing  
in the fluctuated fields,
study their eqs of motion (that are the Einstein, Maxwell and 
Bianchi eqs). The system is reduced to a Schroedinger problem. When 
solved, has eigenfunctions that we identify with the 
glueballs and eigenvalues that are identified with their masses. The 
fluctuations of the fields are dual to different operators in the 
gauge theory and what we are actually computing in the field theory 
side is the two-point correlation function of two glueball 
operators that should behave in a  Wilson expansion as
$$
<O(x) O(y)>= \sum_j c_j e^{M_j |x-y|}
$$
where $M_j$ are the glueball masses. The quantum numbers of the 
glueballs $J^{PC}$ are determined on the basis of the spin ($J$) of 
the supergravity field and the R-symmetry quantum numbers (in a  
KK-harmonics decomposition) as studied, for example, 
in \cite{Kim:1985ez}.  We 
should point that this procedure is not totally clear in many of the 
available confining-models and it should be important to understand 
it better.

This machinery has been applied to some confining models.
Let us add that, since many of the existing Supergravity models are 
duals to confining field theories with only adjoint matter content, 
the objects under study are only glueballs (no hybrids) and since 
we work in the large $N_c$ regime, the glueballs are stable.

Let us briefly review what was done in this subject.
the original idea, described above, has been proposed by 
Witten in \cite{Witten:1998zw}. Many papers followed, exploring this 
nice idea in different contexts. For example, confining models using 
black hole geometries were developed for $QCD_3$ and $QCD_4$ in 
\cite{Csaki:1998qr}, \cite{Brower:1999nj}. Also, models based on rotating 
branes
were introduced \cite{Csaki:1998cb} 
and other models based on 
$AdS_5$ with no SUSY \cite{Constable:1999ch}. All these models have 
an spectrum that is numerically  very close to the one obtained by 
Lattice methods.

We should stress, that even when the comparisons
between the lattice and ``AdS'' based results seem so 
accurate and promising, 
these calculations are done in opposite regimes. Indeed, the 
gravity-dual 
computation is in strong 't Hooft coupling and this limitation is 
imposed by 
the Supergravity approximation. On the other hand, the Lattice 
computations are done at weak coupling, this seems to be a necessity
of having a  continuum limit because the lattice spacing $a$, 
has a 
relation  $a \Lambda_{QCD}\approx e^{-1/g^2 N}$ with the QCD 
coupling and scale (other regularizations give similar results). One might 
think about doing strong coupling lattice computations, but they do 
not seem to be 
smoothly  related to the continuum theory. It is possible that 
the numerical coincidences aluded above, are based on some dynamical 
principle to be understood.

There exist a set of Supergravity models that preserve $N=1$ SUSY 
that have been object of lots of study and amusing advances.
One of the models was put forward by Klebanov and Strassler 
\cite{Klebanov:2000hb} and the glueballs in this model were 
carefully studied in the set of papers  
\cite{Caceres:2000qe},\cite{Gubser:2004qj}.
The results  indicate that, for the Klebanov-Strassler model, the masses of the $0^{++}, \; 1^{--}, \; 
2^{++}$ in the strong 't Hooft coupling limit,  fall in a  linear trajectory. Also, the finding of a 
massless excitation showed that this cascading field theory is not 
in the same universality class of QCD, because of the reasons  
we explained above. \footnote{We thank Oliver Jahn for extensive 
discussions 
on many of the
points touched in this introduction}
\subsection{Motivations and organization of this paper}

We mentioned above a set of Supergravity duals to confining models 
and up 
to this point we just commented on the one proposed by Klebanov and 
Strassler \cite{Klebanov:2000hb}. There exist some other models that 
are based on wrapped D-branes. 
The main idea here is to consider the low energy field theory in 
$(k+1)$ dimensions, obtained by wrapping a Dp brane on a 
$(p-k)$ cycle. Some subtleties of this type of models will be 
explained in  section 2 of this paper. Here we just want to 
emphasize that the glueballs spectrum in this case is poorly understood.

Indeed, there is a paper \cite{Ametller:2003dj}, where a study was 
initiated. We believe that this study is not completely correct from 
a technical viewpoint (we believe that incorrect eqs were used) and 
the 
conclusions expressed there are, even when intuitively 
understandable, also not totally correct. The main point of that paper is 
that 
in one of these models it is necessary to introduce a hard cut-off 
in order to have a discrete spectrum. In this paper 
we re-analize this statement and propose a different result, 
basically that all these models do have discrete spectrum of 
glueballs and give a way of computing it.  
The spectrum even though discrete
is not normalizable. In order to get normalizable states 
we need to introduce a regularization procedure. 
The regularization we propose is not the  
introduction of  a hard cut-off but is more in the spirit
of the Wilson loop calculations 
\cite{Rey:1998ik},\cite{Maldacena:1998im} where a
non-physical part is subtracted. In this  paper we will not make much 
emphasis on the numerical aspects of the problem. Indeed, even when 
discrete  states are numerically obtained, we will not
worry here about comparisons with lattice results, that as 
explained above are perhaps not very significative. The main objective of 
this 
work is to study  qualitative features of the spectrum, point out differences with previously studied cases, and propose a procedure of computing and regularizing in these 
wrapped branes set-ups.

This paper is divided in two parts, one dealing with a 
particular type  IIB model and the other with a type IIA model. Both 
parts have been written and can be read in parallel and almost 
independently.

In section 2, we describe in detail the two models we will be 
using, 
one based in type IIB, with D5 branes wrapping a two-cycle inside 
the resolved conifold. The other in type IIA, based on D6 branes 
wrapping a three cycle in the deformed conifold. Section 3 deals 
with the glueballs in the type IIB set-up, while section 4 
sketches the results corresponding to the  
type IIA model. We did not carry the problem to an end because of 
the need of a more precise numerical analysis, since the solution 
is only numerically known. Section 5 presents conclusions 
and possible future work proposed to the interested reader.
There are very detailed Appendixes that carefully explain all the 
computations in sections 3 and 4.

\section{${\cal N}=1$ SYM models from wrapped branes}
In this section, we write an account of duals to N=1 SYM from 
wrapped branes. The two main models on which we will concentrate are 
the ones based on  D5 branes wrapping  a two cycle, that we will 
consider to be a two-sphere inside a CY3 fold and D6 branes on a  
three 
cycle (a squashed three sphere), also inside a CY3 fold. We will 
present the solutions in detail, and  explain the main 
characteristics of the  dual gauge 
theory. We will emphasize the existence of `extra' modes called KK 
modes with mass of the same order of the confinement scale. Since 
our interest in 
this paper is  on glueballs, we will discuss the influence of 
this `extra' modes in the computation of glueballs for $N=1$ 
SYM.

\subsection{D5 branes wrapping $S^2$}
We will work with the model 
presented in \cite{Maldacena:2000yy}
(the solution was first found in a 
4d context in \cite{Chamseddine:1997nm}) 
and described and studied in more detail in the 
paper \cite{Nunez:2003cf}. Let us briefly describe the 
main points of this supergravity dual to $N=1$ SYM and its UV 
completion.

Suppose that we start with N $D5$ branes, 
the field theory living on them is 
$(5+1)$SYM with 16 supercharges. Then, 
suppose that we wrap two directions of the 
D5 branes on a curved two manifold 
that can be choosen to be a sphere.
In order to preserve SUSY a twisting procedure 
has to be implemented. 
The one we will be interested in this 
section, deals with a twisting that preserves 
four supercharges. In this case 
the two-cycle mentioned above lives inside a CY3 fold. 
Notice that this supergravity solution will be 
dual to a 
four dimensional field theory, only for low energies (small values 
of the 
radial coordinate). 
Indeed, at high energies, the modes of the gauge theory 
start to explore the two cycle and the theory becomes first N=1 SYM 
in six 
dimensions and then, the blowing-up of the dilaton forces us to 
S-dualize and a 
little string theory completes the model in the UV. In this sense, 
to study only the 4d-SYM part of the background, a procedure that 
``substracts'' the unwanted UV completion, should be useful. We will 
elaborate on this in Section 3.1.

The supergravity solution corresponding to the case of interest in 
this 
section, the one preserving four supercharges, has the topology of 
$R^{1,3}\times R\times S^2\times S^3$ and there is a fibration 
between the 
two spheres that allows the SUSY preservation.  The topology of 
the metric, near $r=0$ is $R^{1,6} \times S^3$. The full solution 
and Killing 
spinors are written in 
detail in \cite{Nunez:2003cf}. Let us revise it here for reference.
The metric  in Einstein frame reads,
\beq
ds^2_{10}\,=\,\alpha' g_s N e^{{\phi\over 2}}\,\,\Big[\,
\frac{1}{\alpha' g_s N 
}dx^2_{1,3}\,+\,e^{2h}\,\big(\,d\theta^2+\sin^2\theta 
d\varphi^2\,\big)\,+\,
dr^2\,+\,{1\over 4}\,(w^i-A^i)^2\,\Big]\,\,,
\label{metric}
\eeq
where $\phi$ is the dilaton. The angles
$\theta\in [0,\pi]$ and
$\varphi\in [0,2\pi)$ parametrize a two-sphere. This sphere is 
fibered in the ten 
dimensional metric by the one-forms
$A^i$ $(i=1,2,3)$.
Their expression can be written in terms of a 
function
$a(r)$ and the angles $(\theta,\varphi)$ as follows:
\beq
A^1\,=\,-a(r) d\theta\,,
\,\,\,\,\,\,\,\,\,
A^2\,=\,a(r) \sin\theta d\varphi\,,
\,\,\,\,\,\,\,\,\,
A^3\,=\,- \cos\theta d\varphi\,.
\label{oneform}
\eeq
The   $w^i\,$'s appearing in eq. (\ref{metric}) are the $su(2)$ 
left-invariant 
one-forms,
satisfying
\bea
w^1&=& \cos\psi d\tilde\theta\,+\,\sin\psi\sin\tilde\theta
d\tilde\varphi\,\,,\rc\rc
w^2&=&-\sin\psi d\tilde\theta\,+\,\cos\psi\sin\tilde\theta
d\tilde\varphi\,\,,\rc\rc
w^3&=&d\psi\,+\,\cos\tilde\theta d\tilde\varphi\,\,.
\label{oneformsu2}
\eea
The three angles $\tilde\varphi$, $\tilde\theta$ and $\psi$ take 
values in the
rank $0\le\tilde\varphi< 2\pi$, $0\le\tilde\theta\le\pi$ and
$0\le\psi< 4\pi$. For a metric ansatz such as the one written in
(\ref{metric}) one  obtains a supersymmetric solution when
the functions $a(r)$, $h(r)$ and the dilaton $\phi$ are:
\bea
a(r)&=&{2r\over \sinh 2r}\,\,,\rc\rc
e^{2h}&=&r\coth 2r\,-\,{r^2\over \sinh^2 2r}\,-\,
{1\over 4}\,\,,\rc
e^{-2\phi}&=&e^{-2\phi_0}{2e^h\over \sinh 2r}\,\,,
\label{MNsol}
\eea
where $\phi_0$ is the value of the dilaton at $r=0$. Near the origin 
$r=0$ the 
function
$e^{2h}$ behaves as $e^{2h}\sim r^2$ and the metric is non-singular. 
The solution of 
the
type IIB supergravity includes a
Ramond-Ramond three-form $F_{(3)}$ given by
\beq
\frac{\alpha'}{N}  F_{(3)}\,=\,-{1\over 
4}\,\big(\,w^1-A^1\,\big)\wedge
\big(\,w^2-A^2\,\big)\wedge \big(\,w^3-A^3\,\big)\,+\,{1\over 4}\,\,
\sum_a\,F^a\wedge \big(\,w^a-A^a\,\big)\,\,,
\label{RRthreeform}
\eeq
where $F^a$ is the field strength of the su(2) gauge field $A^a$, 
defined as:
\beq
F^a\,=\,dA^a\,+\,{1\over 2}\epsilon_{abc}\,A^b\wedge A^c\,\,.
\label{fieldstrenght}
\eeq
The different components of $F^a$ are:
                                                                                                    
\beq
F^1\,=\,-a'\,dr\wedge d\theta\,\,,
\,\,\,\,\,\,\,\,\,\,
F^2\,=\,a'\sin\theta dr\wedge d\varphi\,\,,
\,\,\,\,\,\,\,\,\,\,
F^3\,=\,(\,1-a^2\,)\,\sin\theta d\theta\wedge d\varphi\,\,,
\eeq
where the prime denotes derivative with respect to $r$.
Since $dF_{(3)}=0$, one can represent $F_{(3)}$ in terms of a 
two-form potential
$C_{(2)}$ as $F_{(3)}\,=\,dC_{(2)}$. Actually, it is not difficult 
to verify that
$C_{(2)}$ can be taken as:
\bea
\frac{\alpha' C_{(2)}}{ N}&=&{1\over 4}\,\Big[\,\psi\,(\,\sin\theta 
d\theta\wedge 
d\varphi\,-\,
\sin\tilde\theta d\tilde\theta\wedge d\tilde\varphi\,)
\,-\,\cos\theta\cos\tilde\theta d\varphi\wedge 
d\tilde\varphi\,-\rc\rc
&&-a\,(\,d\theta\wedge w^1\,-\,\sin\theta d\varphi\wedge 
w^2\,)\,\Big]\,\,.
\label{RR}
\eea
Moreover, the equation of motion of $F_{(3)}$ in the Einstein frame 
is
$d\Big(\,e^{\phi}\,{}^*F_{(3)}\,\Big)=0$, where $*$ denotes 
Hodge duality. Let us stress here that the previous configuration is 
non-singular. 

Finally, let us comment on the fact that the BPS equations also 
admit a solution in which the function
$a(r)$ vanishes, \ie\ in which the one-form $A^i$ has only one 
non-vanishing
component, namely $A^{3}$. We will refer to this 
solution as the ``abelian'' (or ``singular'') ${\cal N}=1$
background. Its explicit form can be easily obtained by taking the
$r\rightarrow\infty$ limit of 
the functions given in eq. (\ref{MNsol}). Notice that,
indeed $a(r)\rightarrow 0$ as $r\rightarrow\infty$ in  eq. 
(\ref{MNsol}).
Neglecting exponentially suppressed terms, one gets:
\beq
e^{2h}\,=\,r\,-\,{1\over 4}\,\,,
\,\,\,\,\,\,\,\,\,\,\,\,\,\,\,\,\,\,(a=0)\,\,,
\eeq
while $\phi$ can be obtained from the last equation in 
(\ref{MNsol}). The metric
of the abelian background is singular at $r=1/4$ (the position 
of the singularity can
be moved to $r=0$ by a redefinition of the radial coordinate). 
This IR singularity of
the abelian background is removed in 
the non-abelian metric by switching on the $A^1,
A^2$ components of the one-form (\ref{oneform}). 
\subsection{Some analysis of this model}
Let us first summarize the field theory aspects of the dual to the 
gravity solution we will be mainly concerned with. The main 
characteristic is that it contains a 
four dimensional Minkowski space, a radial direction and  a two 
sphere fibered over a three sphere. 
In \cite{Maldacena:2000yy}
this solution was argued to 
be dual to $N=1$ SYM. Let us analize the claim a little more, the 
field theory at low energies (low compared to the inverse size of 
the two-sphere)  has degrees of freedom  given by
a vector field and a Majorana spinor (in 4d). When increasing in 
energy, other modes with mass of the order of the inverse size of 
the $S^2$ appear in the spectrum. These are called KK modes and can 
be seen as coming from the reduction of the maximally D5 branes SUSY 
field 
theory  on a two dimensional sphere and a twisting (explained 
below) are performed. When the energy is high enough the 
excitations of the theory propagate in $5+1$ dimensions and the 
UV-completion of our minimally SUSY four dimensional field theory
is the six dimensional little string theory living on $N$ NS5 
branes. 

Let us recall briefly the twisting procedure. One has a $5+1$ 
field theory (that lives on $N$ D5 branes) that has gauge fields, 
fermions and four scalars, all 
in the adjoint of the $SU(N)$ gauge group. We rewrite the $SO(1,5) 
\times SU(2)_L \times SU(2)_R$ group quantum numbers 
of the fields above in 
terms of 
$SO(1,3) \times SO(2) \times
SU(2)_L \times SU(2)_R$ and then we mix the quantum numbers respect 
to 
$SO(2)$ with those of another $SO(2)$ that lives inside one of the 
$SU(2)'s$. 
After this twisting procedure is performed, we are left with fields 
that under $SO(1,3)\times U(1)\times SU(2)$ transform as 
\beq
A^{a}_{\mu}= (4, 0, 1), \;\; \Phi^a=(1, \pm, 1), \;\; 
\phi^{a}= 2 (1, \pm, 2). \;\;
\label{twistscalar}\eeq
for the  bosons that can be seen to be  a massless gauge field, a 
massive scalar (coming from the gauge field) and other massive 
scalars 
(that originally represented the positions of the D5 branes in 
$R^4$). As a   general rule, all the fields that do transform under 
the 
twisted $U(1)$, the second entry in the charges above, will be 
massive. 
For the fermions we will have 
\beq
\psi^a=(2,0,1),\;(\bar{2},0,1), \;\; (2, ++,1), 
\; (\bar{2},--,1), \;(2,0,2), \; (\bar{2},0,2)
\label{twistfermion}
\eeq
that is a Majorana spinor in four dimensions that is massless and 
then we 
have massive ones (those whose quantum number under the twisted 
$U(1)$ is 
not 
zero). The KK modes are the massive modes mentioned above. Their 
mass is of the order $M_{KK}^2= (R_{S2})^{-2}= \frac{1}{g_s \alpha' 
N} $. 

The dynamics of these KK modes, mixes with the dynamics of 
confinement in this model, because the scale of strong coupling of 
the theory is of the order of the KK mass. If we could work with 
a sigma model for the string in this background (or in the 
S-dual NS5 background) to all orders in $\alpha'$  we could 
decouple both behaviours. Meanwhile, the dynamics of these KK modes 
has not been 
studied in great detail, but some progress have been made, for 
example in the papers \cite{Gimon:2002nr}. Finally, we would like to
mention a paper where a very careful study of the KK modes spectrum have
been done, also pointing a coincidence with $N=1^*$ theory in a  given
Higgs vacuum \cite{Andrews:2005cv}.

Let us briefly comment on the influence of these KK modes
on the glueballs spectrum. Indeed, once the strong coupling regime 
of the field theory is attained, one possible way to compute is using 
these supergravity backgrounds. Given that we are in the supergravity 
approximation, the spectrum of our model includes these KK 
`contaminations' (this feature repeats in all the dual to  
non-conformal field theories). Obviously, our glueballs will be of 
two 
types, those coming from condensates of the gluon and gluino, 
those `composed' out of KK modes and finally, hybrids, composed out of 
SYM fields and KK modes. We would like to discard those with some KK 
constituent. We will comment in the conclusion section on a possibility 
to do this.

Finally, let us mention that there are many succesful checks showing 
that the supergravity background presented above captures different 
non-perturbative aspects of $N=1$ SYM. We will not discuss these 
many checks here, instead, we refer the interested reader to the 
very careful reviews \cite{Bertolini:2003iv}.

\subsection{D6 branes wrapping $S^3$}
Now, let us comment on the models based on D6 branes wrapping a 
calibrated three-cycle inside a CY3 fold.
The progress in this direction originated from the duality between
Chern-Simons gauge theory on $S^3$ at large $N$ and topological 
string
theory on a blown up Calabi-Yau conifold
\cite{Gopakumar:1998ki}. This duality was embedded in string theory 
as
a duality between the IIA string theory of $N$ D6-branes wrapping 
the
blown up $S^3$ of the deformed conifold and IIA string theory on the
small resolution of the conifold with $N$ units of two form
Ramond-Ramond flux through the blown up $S^2$ and no branes
\cite{Vafa:2000wi}. The D6-brane side of the duality involves an
${\mathcal{N}}=1$ gauge theory in four dimensions that is living on
the non-compact directions of the branes, at energies that do not 
probe
the wrapped $S^3$.

Just like before, in
order for the wrapped branes to preserve some supersymmetry, one has 
to
embedd the spin connection of the wrapped cycle into the gauge 
connection,
which is known as twisting the theory.

When we have flat D6 branes, the symmetry group of the 
configuration is $SO(1,6)\times SO(3)_R$. The spinors transform in 
the
{\bf (8,2)} of the isometry group and the scalars in the
{\bf (1,3)}, whilst the gauge particles are in the {\bf (7,1)}
\cite{Seiberg:1997ax}.
Wrapping the D6 brane on the three-sphere breaks the group to 
$SO(1,3)\times 
SO(3)\times SO(3)_R$. The technical meaning of twisting is that the 
two $SO(3)$s get mixed to allow the existence of four dimensional
spinors that transform as scalars under the new twisted $SO(3)$ 
\cite{Edelstein:2001pu}. One
can then see that the remaining particles in the 
spectrum that transform as scalars under the twisted $SO(3)$ are the 
gauge
field and four of the initial sixteen spinors. Thus the 
massless field 
content is that 
of ${\mathcal{N}}=1$ SYM. Like in the model analyzed in the 
previous section, apart from these fields, there will be 
massive modes, whose mass
scale is set by the size of the curved cycle. When we probe the 
system 
with very low energies, we find only the spectrum of 
${\mathcal{N}}=1$ SYM. 
For D6 branes in flat space, the `decoupling' limit does 
not completely
decouple the gauge theory modes from bulk modes 
\cite{Itzhaki:1998dd}. In our case, we expect a 
good gauge theory description only when the size of the wrapped 
three-cycle is 
large, which implies that we have to probe the system with 
very low energies to get 3+1 dimensional SYM \cite{Atiyah:2000zz}. 
In this case, the size of the two cycle
in the flopped geometry is very near to zero, so a good gravity 
description 
is not expected. In short, we must keep in mind that the field 
theory we will be 
dealing with has more degrees of freedom than pure ${\mathcal{N}}=1$ 
SYM, thus the glueballs masses that one might obtain following 
the procedure explained in the following sections might be 
`contamined' by glueballs composed out of KK modes or hybrids 
composed out of KK modes and gluons or gauginos. Again, how to 
decouple the ones we are interested into from those 
glueballs `composed' of KK modes is going to be discussed in the 
conclusion section.

Finally, let us add that
the duality described above is naturally
understood by considering
M-theory on a $G_2$ holonomy metric \cite{Atiyah:2000zz}. In eleven
dimensions, $G_2$ holonomy implements ${\mathcal{N}}=1$ as pure 
gravity.
One starts with a singular $G_2$ manifold that on dimensional
reduction to IIA string theory corresponds to $N$ D6 branes wrapping
the $S^3$ of the deformed conifold. There is an $SU(N)$ gauge theory
at the singular locus/D6 brane. This configuration describes
the UV of the gauge theory. As the coupling runs to the IR, a blown 
up
$S^3$ in the $G_2$ manifold shrinks and another has fixed size. This flop
is smooth in M-theory physics. The metrics will be discussed in more
detail in the following sections. In the IR regime, the $G_2$
manifold is non-singular and dimensional reduction to IIA gives
precisely the aforementioned small resolution of the conifold with 
no
branes and RR flux. 

Let us now, write explicitly the background on which we will be 
interested.
It is conveninet to start with the eleven dimensional M-theory 
background, that reads
\beq
ds_{11}^2= dx_{1,3}^2+ ds_7^2
\eeq
with
\bea
ds^2_7 = dr^2 + a(r)^2\left[(\S_1 + g(r)\s_1)^2
+(\S_2 + g(r)\s_2)^2\right]+ c(r)^2(\S_3 + g_3(r) \s_3)^2  \nn \\
+ b(r)^2\left[\s_1^2 + \s_2^2 \right] + f(r)^2 \s_3^2 ,
\eea
where $\S_i,\s_i$ are left-invariant one-forms on the $SU(2)$s 
(\ref{oneformsu2}).
The six functions are not all independent
\be
g(r) = \frac{-a(r)f(r)}{2b(r)c(r)} , \;\; g_3(r) = -1 + 2g(r)^2 .
\ee
None of the radial functions are known explicitly, although the
asymptotics at the origin and at infinity are known. The asymptotics
are found by finding Taylor series solutions to the first order 
equations
for the radial functions. The equations are 
\cite{Cvetic:2001ih},\cite{Brandhuber:2001kq}
\bea\label{eq:Deqns}
\dot{a} = -\frac{c}{2a} + \frac{a^5 f^2}{8 b^4 c^3}, & \;\; &
\dot{b} = -\frac{c}{2b} - \frac{a^2 (a^2-3c^2)f^2}{8b^3c^3}, \nn \\
\dot{c} = -1+\frac{c^2}{2 a^2}+\frac{c^2}{2 b^2}-\frac{3 a^2
f^2}{8b^4}, & \;\; &
\dot{f} = -\frac{a^4 f^3}{4 b^4 c^3}.
\eea
As $r\to 0$ one has
\bea\label{eq:Dat0}
a(r) & = & \frac{r}{2}-\frac{(q_0^2+2)r^3}{288 R_0^2} -
\frac{(-74-29q_0^2+31q_0^4)r^5}{69120 R_0^4} + \cdots  , \nn \\
b(r) & = & R_0 - \frac{(q_0^2-2)r^2}{16 R_0} -
\frac{(13-21q_0^2+11q_0^4) r^4}{1152 R_0^3} + \cdots  , \nn \\
c(r) & = & -\frac{r}{2} - \frac{(5q_0^2-8)r^3}{288 R_0^2} -
\frac{(232-353q_0^2+157q_0^4) r^5}{34560 R_0^4}+ \cdots , \nn \\
f(r) & = & q_0 R_0 + \frac{q_0^3 r^2}{16 R_0} +
\frac{q_0^3(-14+11q_0^2) r^4}{1152 R_0^3} + \cdots ,
\eea
where $q_0$ and $R_0$ are constants.
Note that $a(r)$ and $c(r)$ collapse and the other two functions do 
not.
As $r\to\infty$ we have
\bea\label{eq:Datinfinity}
a(r) & = &  \frac{r}{\sqrt{6}} - \frac{\sqrt{3} q_1 R_1}{\sqrt{2}} +
\frac{(27\sqrt{6}- 96 h_1 ) R_1^2}{96r} + \cdots , \nn \\
b(r) & = &  \frac{r}{\sqrt{6}} - \frac{\sqrt{3} q_1 R_1}{\sqrt{2}} +
\frac{h_1 R_1^2}{r} + \cdots , \nn \\
c(r) & = &  \frac{-r}{3} + q_1 R_1 - \frac{9 R_1^2}{8r} + \cdots , 
\nn \\
f(r) & = &  R_1 - \frac{27 R_1^3}{8 r^2} - \frac{81 R_1^4 q_1}{4 
r^3} + \cdots .
\eea
With constants $R_1,q_1,h_1$. Note that $f(r)$ stabilises.
Three constants appear to this order, whilst there were only two
constants in the expansion around the origin. This just means that 
for
some values of these constants, the corresponding solution will
diverge before it reaches zero. In any case, we find no $h_1$
dependence in the results below.

We can reduce this to Type IIa and we will find a non-singular 
background with dilaton, metric and RR one form excited, that reads,
\bea
& & ds_{IIA,string}^2= 2 e^{2/3\phi}\Big( dx_{1,3}^2 + dr^2 + b(r)^2 
(\s_1^2+\s_2^2) +a(r)^2 ((\S_1+ g(r)\s_1)^2+(\S_2 + g(r)\s_2)^2) 
+\nonumber\\ & &\frac{f^2c^2}{f^2 +c^2(1+g_3)^2}(\S_3 -\s_3)^2   
\Big)\nonumber\\
& & 4 e^{4/3\phi}= f(r)^2 + c(r)^2 (1+ g_3(r))^2, \; 
A_1=\cos\theta d\varphi +\cos\tilde{\theta}d\tilde{\varphi} 
+\frac{f^2 - c^2(1-g_3^2)}{f^2+ c^2(1+g_3)^2}(\s_3 - \S_3)
\label{conf2a}
\eea
Where we have defined $dx_{11}= d\psi'+d\psi$ and $d\hat{\psi}= 
d\psi - d\psi'$.
So, to summarize the things clearly, let us write
the metric of our  IIA solution in Einstein frame (as will be 
used below), 
\bea
& & ds_{E}^2= 2 e^{\phi/6}\Big( dx_{1,3}^2 + dr^2 + ( b(r)^2 + 
g(r)^2) (d\theta^2 + \sin^2\theta d\varphi^2) +a(r)^2 
(d\tilde{\theta}^2 + \sin^2\tilde{\theta} d\tilde{\varphi}^2)
+\nonumber\\ 
& & + 2 g(r) a(r)^2 [\cos\hat{\psi} (d\theta d\tilde{\theta} + 
\sin\theta \sin\tilde{\theta} d\varphi d\tilde{\varphi}) 
+\sin\hat{\psi}(\sin\theta 
d\tilde{\theta} d\varphi - sin\tilde{\theta} d\theta 
d\tilde{\varphi})] +\nonumber\\
& &
+\frac{f^2c^2}{f^2 +c^2(1+g_3)^2}(d\hat{\psi} + \cos\theta 
d\varphi - \cos\tilde{\theta} d\tilde{\varphi})^2
\Big)
\label{meteisnt2ap}
\eea
with the same dilaton as in (\ref{conf2a}), besides, the field 
strength $F_2$ reads,
\beq
F_2= k'(r) dr \wedge (d\hat{\psi} +\cos\theta d\varphi - 
\cos\tilde{\theta}d\tilde{\varphi}) - (k(r) +1) \sin\theta d\theta 
\wedge d\varphi + (k(r)-1) \sin\tilde{\theta} d\tilde{\theta}\wedge 
d\tilde{\varphi}, 
\eeq
and $$ k(r)= \frac{f^2 - c^2(1-g_3^2)}{f^2+ c^2(1+g_3)^2}.$$

To end this section, let us briefly revise what checks exist of the 
duality between the backgrounds presented here and $N=1$ SYM.

Of course, the number of supercharges match, there is a nice picture 
of confinement in terms of a Wilson loop computation in IIA. But
most of the presently known matchings of
$N=1$ SYM with $G_2$ holonomy M-theory come from considering 
membrane
instantons as gauge theory instantons that generate the
superpotential \cite{Acharya:1998pm}, membranes wrapped on 
one-cycles in the IR
geometry that are super QCD strings in the gauge theory 
\cite{Acharya:2000gb,Acharya:2001hq}, and fivebranes
wrapped on three-cycles that give domain walls in the gauge theory 
\cite{Acharya:2000gb,Acharya:2001dz}.
These matchings above, are essentially
topological and do not use the explicit form of the $G_2$ metrics. 
In the category of test/checks that use the form of the metric, we 
can mention \cite{Hartnoll:2002th}, where rotating membranes in 
these $G_2 $ backgrounds have been studied and relations for large 
operators in SYM have been reproduced. We should also mention 
\cite{Gursoy:2003hf}, where a very nice picture of the chiral 
anomaly of SYM have been developed. Perhaps less promising is the 
fact that confining string tensions do not arise as cleanly in the IIA 
backgrounds as in the type IIB case \cite{Hartnoll:2004yr}. 
There are many 
aspects of 
this duality that are not on a very firm basis and we think that through 
study, these unclear points might become clear. The results that 
we will present in Section 4, use the 
explicit form of the 
metric and should be considered to belong to this second category 
of tests.

\section{Glueballs from type IIB solution}
As we explained above, in order to study glueballs, we need the 
variation of the eqs of motion.
In the type IIB case, for the solution of D5 branes wrapping $S^2$ 
inside a $CY_3$-fold discussed in the previous section, the Einstein 
eqs for the metric, dilaton and three form read, \footnote{We will 
denote the contraction of indexes with $*$ symbols, so, for 
example $g^{\mu\nu}A_{\mu kl} B_{\nu jp}= A_{\mu kl} B^{\mu}_{jp}= 
A_{*kl} B^{*}_{jp}$.} 
\beq
R_{\mu\nu}=\frac{1}{2}\partial_\mu\phi\partial_\nu\phi 
+\frac{g_s e^{\phi}}{4}\Big( F_{\mu**} F_\nu^{**} - 
\frac{1}{12}g_{\mu\nu}F_3^2   \Big)
\label{riccieq2b}
\eeq
by contracting we get the Ricci scalar eq, 
\beq
R=\frac{1}{2}(\partial_\mu \phi)^2 + \frac{g_s e^\phi}{24}F_3^2 .
\label{ricciscalareq2b}
\eeq
The dilaton, Maxwell and Bianchi eqs. are,
\beq
\nabla^2\phi= \frac{g_s e^\phi}{12}F_3^2,\;\;\partial_\mu\Big( 
\sqrt{g} e^\phi F^{\mu\nu\rho} \Big)=0,\;\; 
\partial_{[\mu} F_{\kappa \nu\rho]}=0
\label{dilmax2b}
\eeq

Now, let us study the fluctuations of these eqs.
Let us assume that the background fields vary according to
\beq
g_{\mu\nu}\to g_{\mu\nu}+ \epsilon h_{\mu\nu},\;\; \phi\to 
\phi+\epsilon\delta\phi, \;\;F_{\mu\nu\rho}\to 
F_{\mu\nu\rho}+\epsilon \delta F_{\mu\nu\rho}
\label{fluct2b}
\eeq
Keeping only linear order in the parameter $\epsilon$, we get 
eqs  for the fluctuated fields (that 
can be found written in detail in Appendix B). The metric fluctuation, $h_{\mu\nu}$
can be splitted in its worldvolume, internal and 
mixed parts: $h_{ij},h_{\alpha\beta}$ and $h_{i\alpha}$
respectively. We assume that there is no fluctuation 
in the mixed part, $h_{i\alpha}=0$ and 
perform a Weyl  
shift \cite{Kim:1985ez} in the worldvolume fluctuation, \footnote { The standard Weyl shift in a 
D dimensional spacetime is $\lambda/5 = -1/(D-2) $; 
this value is required to simplify the variation 
of $\nabla^2$. Here we choose to leave 
$\lambda$ as a constant to be determined later.}  
\bea
& & h_{i j} = h'_{ij}+\frac{\lambda}{5} g_{i j} h^\alpha 
_\alpha,\;\;,\;\;
h_{\alpha \beta } = h_{(\alpha \beta)} + \frac{1}{5} g_{\alpha \beta } 
h^{a}_{a}
\label{shifts}
\eea
We denote with latin indices the worldvolume and transverse coordinates  $i,j=\vec{x},t,r$  and with with greek indices the internal coordinates, $\alpha,\beta=\theta,\tilde{\theta},\varphi,
\tilde{\varphi}, \psi$. Also,   $\lambda$ is a constant and $h_{(\alpha\beta)}$ is the symmetric 
traceless part of the fluctuation in the internal directions. 
Notice that  the trace of the metric fluctuation over all the space 
is
$h^\mu_\mu = (\lambda +1)  h^\alpha _\alpha $. From here on we will denote 
with $x$ 
the radial and `gauge theory' coordinates and with $y$ the internal 
coordinates (the angles on the spheres).
Imposing a de Donder and Lorentz type condition, $\nabla^\alpha 
h_{(\alpha\beta)}= \nabla^\alpha h_{\alpha i} =0$,  the  decomposition in 
harmonics for the fluctuated fields is, 
\bea
h^\alpha_\alpha (x,y) =\Sigma_I \Pi^I (x) Y^I (y),\;\;
h_{(\alpha\beta)} =\Sigma_I b^I(x)Y^I_{(\alpha\beta)}(y),\nonumber\\  
h_{ij}=\Sigma_I H_{ij}(x)Y^I(y),\;\;\delta\phi (x,y)= \Sigma_I f^I (x) Y^I (y) \nonumber\\
 C_{ij} =\Sigma_I a_{ij}^I(x)Y^I (y),\ \ C_{i\alpha}=\Sigma_I a_i^I (x) Y_\alpha^I(y) ,\ \  C_{\alpha\beta}=\Sigma_I a^I (x) Y_{\alpha \beta}^I(x).
\label{fluctarmonicsb}
\eea
Where $C_2$ is the two form potential. We set $h'_{ij} =0$. Given this ansatz the equations of motion can be consistently solved for the other fluctuations. 
Using the decomposition in harmonics, keeping only the s-wave and choosing a particular value for $\lambda= 
-\frac{35}{29}$ the two equations for 
the fluctuations nicely combine into just one equation (again, see the 
Appendix B for details). This final eq. reads,
\eqn{final}{\nabla^2 f(x)+ 2g^{rr}\partial_r f(x)
\partial_r \phi +\frac{g_s}{12} e^\phi
\Big[ \frac{5}{2}F_3^2 +\frac{29\cdot 3}{10}
(g^{ab}F_{a**}F_b^{**} -\frac{35}{29}g^{rr}F_{r**}F_r^{**}) 
\Big]f(x) =0}

Expanding (\ref{final}) in plane waves, $f(x) = F(r) e^{(i K\cdot x)}$, we have,
\bea
g^{rr}\frac{d^2 F(r)}{dr^2} &+&\left(2g^{rr}\partial_r \phi + \frac{1}{\sqrt g} \partial_r ({\sqrt g}g^{rr})\right)\frac{d F(r)}{dr} \nonumber\\
 &+&\left( \frac{g_s}{12} e^\phi
[ \frac{5}{2}F_3^2 +\frac{29\cdot 3}{10}
(g^{ab}F_{a**}F_b^{**} -\frac{35}{29}g^{rr}F_{r**}F_r^{**})] -K^2 g^{xx}\right)F(r)=0\nonumber\\
\label{finalplane}
\eea

\begin{figure}
   \centerline{\epsfig{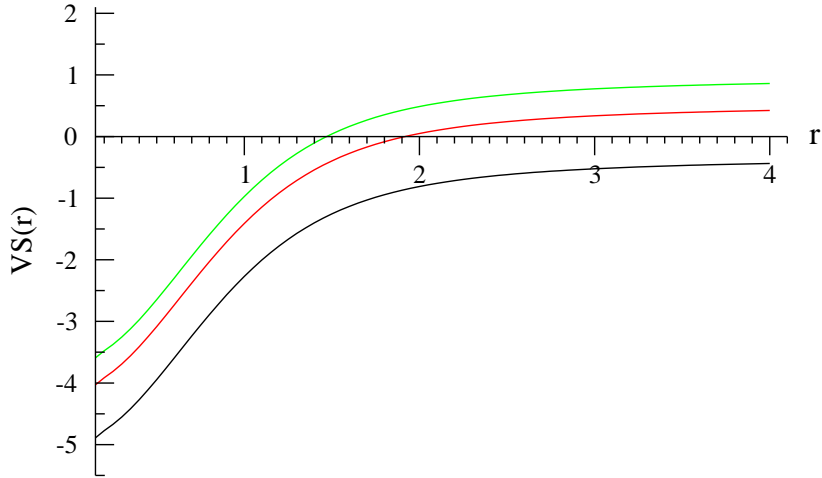}}
   \caption{\label{fig:IIBpotential}VS(r) for different values of $K^2$.}
\end{figure}

This  equation  will be  solved  numerically;  
the glueball masses are given by the eigenvalues
$K^2$ for which there is a solution with appropriate 
boundary conditions. Before studying the 
boundary conditions of this problem let us cast eq. 
(\ref{finalplane}) in a more familiar way. 
To save us some writing, denote  the coefficient of 
the first derivative term ($\frac{dF}{dr}$) in (\ref{finalplane}) 
as $\mu(r)$  and the coefficient of $F(r)$ as $\alpha(r)$, that is,
\bea
&&\mu(r) =2\partial_r \phi + \frac{g_{rr}}{\sqrt g} \partial_r ({\sqrt g}g^{rr}) \nonumber \\
&&\alpha(r)= \frac{g_s g_{rr}}{12} e^\phi
[ \frac{5}{2}F_3^2 
+\frac{29\cdot 3}{10}(g^{ab}F_{a**}F_b^{**} 
-\frac{35}{29}g^{rr}F_{r**}F_r^{**})]
\eea
Making a change of  variables, 
$\cF (r) = e^{-\frac{1}{2}\int ^r \mu(z) dz} F(r), $ equation 
(\ref{finalplane}) can be written in a Schr\"odinger form
\eqn{Scheq}{\frac{d^2 \cF(r)}{dr^2 } - VS(r) \cF =0,}
where
\eqn{V1}{VS(r)= -\alpha(r) +\frac{1}{4}(\mu^2(r) + 2 
\frac{d\mu(r)}{dr}) +K^2 }
A graph of the  Schr\"odinger potential VS(r) is given in Figure 
\ref{fig:IIBpotential}. 
At this point it is convenient to recall that, as explained in section 
two,  
the full model, considering its UV completion  is not dual to 
a non-abelian gauge theory, but to a little string theory. 
Indeed, in the UV, due to the divergent dilaton, the solution has to be 
S-dualized 
yielding a IIB solution with NS five branes. 
In the decoupling limit this 
background   
is not dual to $\cN =1$ SYM in four dimensions but to a higher 
dimensional, 5+1, little string theory. Naturally,  trying to calculate 
observables  by simply using the solution up to infinity  might not yield 
 sensible answers. In what follows  we want to study the glueball spectrum and, if necessary, propose a regularization procedure.

To calculate the  $0^{++}$ mass we have to 
numerically find  eigenvalues satisfying equation and appropriate boundary 
conditions. The asymptotic behavior of the potential is, 
\eqn{Vinf}{VS(r \rightarrow \infty )= -\frac{92}{5} + K^2}
\eqn{Vzero}{VS(r\rightarrow 0) = -\frac{1036}{45} + K^2}
Choosing  the exponentially decreasing solution at  infinity we get, 
\eqn{bcinf}{\cF(r\rightarrow \infty) \sim  e^{-\sqrt{ -\frac{92}{5} +K^2 } 
r } .} 
At the origin we demand a smooth solution,
\eqn{bc0}{\frac{d F(r)}{dr }=0 .}
Similarly to Klebanov-Strassler, satisfying the boundary conditions implies 
that the eigenvalues are bounded from below, $ K^2 > 92/5$ and thus, there 
is a mass gap. But here, in addition of being bounded from below,   the 
eigenvalues are also bounded from above,  $K^2< 1036/45 $. 
A similar phenomenon was observed in \cite{Arean:2005ar} 
in the context of non-commutative gauge theories.

Another difference with  the Klebanov-Strassler model is that  here the 
boundary condition at infinity depends on the eigenvalue. In a technical 
sense, each eigenvalue defines a different  problem -different boundary 
condition- and the spectrum is then  given by the eigenvalues of this 
collection of problems; It is a more general situation than the standard 
eigenvalue problem. 
   
Using the  WKB method \cite{Danielsson:1998wt}
we can estimate the 
eigenvalues $K^2$ for which there 
exists a 
solution of (\ref{Scheq}) satisfying (\ref{bcinf}) and(\ref{bc0}). Also, it can be shown numerically that the WKB integral $\int _0^ {r({\small K^2})}{ \sqrt{-VS(r) }} dr $ is a monotonically decreasing function of $K^2$. And this fact can be used to prove that there is only one eigenvalue in the spectrum. We find,   
$K_0= 4.33 \ (1/\sqrt {g_s\alpha' N})$.  However, it is easy to show that  
this eigenvalue does not correspond to a normalizable state. For 
large r, 
$\sqrt{ {\rm det} \  g}\sim e^{\frac{5}{2} r}$ and thus,
\eqn{esta}{ \sqrt{ {\rm det} g} \cF_0 ^2 \sim  e^{(\frac{5}{2}  - 
2\sqrt{(-\frac{92}{5} +4.3^2)} r})}
\noindent so the integral $\int_0^\infty \sqrt{{\rm det} g} \cF_0^2 $ does 
not converge. The issue we are confronted with now is to find a good 
regularization for this model.

Let us note an important point. Sean Hartnoll pointed that one 
might think of 
taking the norm in flat space $\int dr |{\cal F}|^2 $. Indeed, the 
equation (\ref{Scheq}) seems to indicate that, but we used a norm 
obtained from 
on the ten dimensional curved background $\int d^{10}x  |{\cal 
F}|^2$, and is this norm the one that is forcing us to some 
regularization. His comment is based on the fact that one should 
only worry about our fluctuations to have finite energy and 
according to the paper \cite{Gibbons:2002pq}, this condition implies 
the finitness of the norm in flat space. If we use his proposal, our 
wave functions ${\cal F}(r)$ are normalizable. In this paper we 
choose to attach to the more conventional norm defined in the 
curved space. The differences and physical implications of each 
choice of norm will be investigated elsewhere.\footnote{We thank 
Sean Hartnoll and Stathis Tompaidis for valuable dicussions and input regarding this 
paragraph.}
 
Coming back to the regularization we need to introduce, let 
us recall the reader 
that, as explained in section 
two, in addition to the solution 
given by (\ref{metric})-(\ref{RR})  there is another solution to 
the BPS equations, that we aluded to as the "abelian" solution. The 
abelian (singular)  and non-abelian (non-singular) 
solutions are the same at infinity but the abelian solution is singular in 
the IR and thus is not a good dual to SYM, that has nothing singular. 
We propose that a 
good regularization for this model is to use 
the background presented in  (\ref{metric})-(\ref{RR}) 
up to the point where it becomes indistinguishable from the abelian one. 
After this point the two solutions are the same and neither of them is dual 
to SYM, but to a higher dimensional theory on which we are not interested 
here. 
 
\begin{figure}
   \centerline{\epsfig{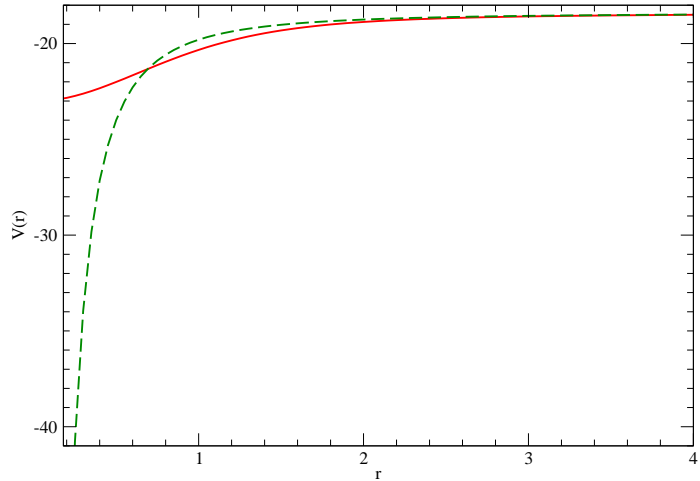}}
   \caption{\label{fig:abvsnonab}VS(r) for abelian and non-abelian 
theories, dashed line and solid lines.}
\end{figure}

Figure (\ref{fig:abvsnonab}) shows a 
plot of the Schr\"odinger potential for the abelian and non abelian 
solutions. Our proposal is that the IR solution that captures  the physics of 
N=1 SYM is valid only up to the region $\Lambda_{Ab}$ where the potential 
becomes the same as the one of the abelian solution. The scale 
$\Lambda_{Ab}$, measured in units of  $g_s\alpha'N$,  is set by the vacuum 
expectation value of the dilaton $\Phi_0$. 

Numerically we are not doing anything new;  the choice of  the right endpoint of integration is always arbitrary, decided to best fit the physical problem at hand.  
Using a generalized shooting technique and integrating up to  the point 
where $V_{Ab} =V_{NonAb}  \pm 0.00001$, we find an improved value for the 
WKB estimate, $K_0 = 4.291 (1/\sqrt {g_s\alpha' N}) $. 
This is a numerically  stable eigenvalue meaning that 
small changes in the initital guess or pushing the 
endpoint of integration further to the right do not afect the value 
obtained.  But  
this eigenstate is not normalizable, to obtain a normalizable state 
we have to impose the regularization procedure proposed above. 
This will be  explained in detail in  the next section.  

It is worth emphasizing that this model produces a 
discrete spectrum even without any kind of regularization. 

\subsection{Understanding The Regularization Procedure}
Above we have proposed that the correct way of computing in this model
is to do a computation with the non-singular solution and, for large 
values of the radial coordinate, substract the 
result obtained with the singular background. This, we proposed is 
calculating in the dual N=1 SYM theory.

Let us get a better intuition of this sort of ``regularization 
procedure''. In physical 
terms, this procedure is easy to 
understand and is just instructing us to 
do our computations only in the 
region that is of interest to $N=1$ SYM.

Indeed, since the non-abelian (non-singular) solution (that captures the 
IR effects of the dual field theory) asymptotes to the 
abelian (singular) solution (that is dual to a higher dimensional field 
theory), what we are basically doing when explicitely 
computing is substract the result obtained with the non-singular 
background minus those obtained at large values of the radial coordinate. 
Basically, this boils down to computing only in the region that is dual to 
$N=1$ SYM (with KK impurities as explained above). This is not very 
different from the type of regularizations done, for example in the 
computation of Wilson loops \cite{Rey:1998ik},\cite{Maldacena:1998im},
where the infinite mass of the non-dynamical quark was substracted, or, 
what is the same, the mass of an infinite string not feeling the effects 
of the background (to which the real string asymptotes) is substracted.
This sort of regularization was also used in \cite{Nunez:2003cf}. In that 
paper, even when a hard cut-off was imposed for numerical convenience, one 
should think about it as the fact that the computation was done in the 
region of interest, where the probe brane that adds flavor to the quenched 
version of N=1 SQCD is 
different from the probe brane in the singular background. More recently
this sort of regularization was used in \cite{Evans:2005ip}, even 
when the model used in that paper is different from ours, we 
believe their regularization can be understood in the lines we wrote 
above.

Some readers might object the following: if one computes in this way, 
everything will give a finite result, so in this case, all functions will 
be normalizable. This questioning is 
valid, so let us try to answer it; for this it will be convenient to 
resort 
on an example where an exact solution is known. 

Hence, it is instructive 
to analyze the solution studied in the paper \cite{Gubser:2004qj}. 
Indeed, 
in that paper, the authors realized that a fluctuation given by
\beq
\delta g_{\mu\nu}=\delta\phi=0, \;\;\; \delta F_3 . F_3=0, \;\; \delta 
F_3= *_4 dA_2
\label{gkh}
\eeq
with no restrictions to the functional form of the two form, solves the 
eqs of motion, that are written in the Appendix B
(\ref{riccieq2ba})-(\ref{delatmaxwell}). 
This could be a massless glueball, but when computing the norm $$\int 
d^{10}x \sqrt{g} |\delta F_3|^2 ,$$  seems to diverge, thus ruling it out 
as an state in the strong coupling theory. 

If we apply our criteria to this case, one might worry that the norm 
computed above will give a finite result, thus leading to a massless 
glueball that one does not expect in this theory (contrary to the KS 
case \cite{Gubser:2004qj}). If these quantities  give a finite result, 
this will imply that an effect not expected (a massless glueball) shows up.

So, to understand this, let us do the computation for the norm, and apply 
our regularization procedure
\beq
||\delta F_3||\propto \int dr e^{2h + 2\phi}
\label{norma}
\eeq
the regularization proposed above, indicates that we do a computation like 
this
\beq
||\delta F_3||=  \int_{0}^{r_0} dr e^{2\phi + 2 
h}|_{non-singular} + \int_{r_0}^{\infty} dr e^{2\phi + 2 
h}|_{non-singular} -  \int_{r_0}^{\infty} dr e^{2\phi + 2 h}|_{singular}
\label{regularizada}
\eeq
Where $r_0$ is a value of the radial coordinate where the functions 
$e^{2h+2\phi}$ computed in the singular and non-singular backgrounds 
are very similar to some degree of precision that is arbitrarily 
fixed. Since both integrands have the same asymptotics, they should equally  
diverge at large values of the radial coordinate. Indeed, both integrals 
diverge at leading order in the same way, but 
contrary to what one might expect, the integrals differ 
in a divergent quantity (and many convergent terms), thus, the computation 
in (\ref{regularizada}) is 
divergent and the configuration in (\ref{gkh}) is not a good state of 
our theory.

It is important to observe that many of the test that the solution has 
passed (see the review articles \cite{Bertolini:2003iv} ), still work with 
this regularization.

We would like to stress that even what was done in \cite{Ametller:2003dj} 
was not technically correct (as we mentioned, they seem to have used the 
wrong fluctuated eqs), the hard cut-off that they introduced is 
doing the same job that the regularization that we proposed here. 
Nevertheless, we have to make clear some important differences with  
\cite{Ametller:2003dj} (apart from the fact that we use different 
eqs.). In  \cite{Ametller:2003dj} the authors did not find a 
discrete spectrum before imposing their hard cut-off, while we found 
one before our regularization, notice that our potential $VS(r)$ 
does `confine' wavefunctions. We need to appeal to the 
regularization, only to satisfy the normalizability condition of our 
discrete states (and this is because we are being conservative and 
adopting a curved space measure for our normalizations). So, the hard 
cut-off regulation is quite different 
from what we have done here. If one introduces a cut-off, together 
with some boundary conditions, all solutions to the Schroedinger eqs 
will be normalizable. In our case, things are more subtle, as we 
explained above.
\section{Glueballs from type IIA perspective}
This sections uses the same methods developed in Section 3 (See 
Appendix B.2 for all the details) for the type IIA background 
explained in Section 2. We will not carry out a full numerical 
analysis like in Section 3, but we will leave the system set for 
this more complicated (fully numerical) problem.

Let us study glueballs for the case of wrapped branes in type IIA 
string theory.
As explained above, in this case, the relevant background consists in $N$ 
D6 branes 
wrapping a three cycle inside a CY3 fold.
So, to summarize the things clearly, let us write
the dilaton and the metric of the IIA solution in Einstein frame, 
\bea
& & ds_{E}^2= 2 e^{\phi/6}\Big( dx_{1,3}^2 + dr^2 + ( b(r)^2 + 
g(r)^2) (d\theta^2 + \sin^2\theta d\varphi^2) +a(r)^2 
(d\tilde{\theta}^2 + \sin^2\tilde{\theta} d\tilde{\varphi}^2)
+\nonumber\\ 
& & + 2 g(r) a(r)^2 [\cos\hat{\psi} (d\theta d\tilde{\theta} + 
\sin\theta \sin\tilde{\theta} d\varphi d\tilde{\varphi}) 
+\sin\hat{\psi}(\sin\theta 
d\tilde{\theta} d\varphi - sin\tilde{\theta} d\theta 
d\tilde{\varphi})] +\nonumber\\
& &
+\frac{f^2c^2}{f^2 +c^2(1+g_3)^2}(d\hat{\psi} + \cos\theta 
d\varphi - \cos\tilde{\theta} d\tilde{\varphi})^2
\Big), \nonumber\\
& & 4 e^{4/3\phi}= f(r)^2 + c(r)^2 (1+ g_3(r))^2.
\label{meteisnt2a}
\eea
The field 
strength $F_2$ reads,
\beq
F_2= k'(r) dr \wedge (d\hat{\psi} +\cos\theta d\varphi - 
\cos\tilde{\theta}d\tilde{\varphi}) - (k(r) +1) \sin\theta d\theta 
\wedge d\varphi + (k(r)-1) \sin\tilde{\theta} d\tilde{\theta}\wedge 
d\tilde{\varphi}, 
\eeq
with $$ k(r)= \frac{f^2 - c^2(1-g_3^2)}{f^2+ c^2(1+g_3)^2}.$$
In the following, we will work with this IIA set-up and because of the 
many similarities 
with the IIB model studied in the previous section, we will use 
the same approach. It is interesting 
to mention that if we can find glueballs in IIA, they should have an 
expression in M theory purely in terms of a metric fluctuation.
Let us start by finding the dynamics of the fluctuations.
The Lagrangian of IIA is \footnote{notice 
that we take $g_s=1$ in this section}
\bea
& & L=\sqrt{g}\Big(R-\frac{1}{2}(\partial\phi)^2 -\frac{1}{4} 
e^{3/2\phi} F_{2}^2 -\frac{1}{12}e^{-\phi}H_3^2 
-\frac{1}{48}e^{\phi/2}\hat{F}_4^2   \Big) + \frac{1}{2}B_2\wedge 
F_4\wedge F_4, \nonumber\\
& & \hat{F}_4= dC_3 - H_3\wedge A_1,\;\; F_2= dA_1, \;\; H_3= dB_2,
\label{2a}
\eea
now, let us focus on the configurations that are of our interest, 
that is those where only the fields $\phi,\;g_{\mu\nu},\; A_\mu$ 
are turned on. The eqs of motion in this case are 
\bea
& & R_{\mu\nu}= \frac{1}{2}\partial_\mu\phi \partial_\nu\phi + 
\frac{1}{2}e^{3/2\phi}(F_{\mu *} F_\nu^* 
-\frac{1}{16}g_{\mu\nu}F^2),\nonumber\\
& & R= \frac{1}{2}(\partial\phi)^2 +\frac{3}{16}e^{3/2\phi}F^2,\;\; 
\nabla^2\phi = \frac{3}{8} e^{3/2\phi}F^2,\nonumber\\
& & \partial_\mu\Big(\sqrt{g} e^{3/2\phi}F^{\mu\nu}   
\Big)=0,\;\;\partial_{[\mu}F_{\nu\rho]}=0
\label{ecsmov}
\eea
Now, let us study the variations of these eqs. Under a fluctuation 
in all the relevant fields
\bea
& & \phi =\phi +\epsilon\delta\phi,\;\;\; g_{\mu\nu}=g_{\mu\nu} 
+\epsilon h_{\mu\nu}\;\;\; F_{\mu\nu}= F_{\mu\nu} + \epsilon \delta 
F_{\mu\nu}.
\eea
Again, we will propose a particular form for the metric fluctuation
\beq
h_{ij}=\frac{\lambda}{5} h g_{ij}, (i,j=x,t,r) \;\;\; 
h_{\alpha\beta}= h_{(\alpha\beta)} +\frac{h}{5} 
g_{\alpha\beta} . 
\label{metricfluct}
\eeq
The same comments that we made regarding this decomposition, before 
eq. (\ref{shifts}) are also 
pertinent here. We have denoted the trace of the internal part of 
the metric as 
$h=h_a^a$. We also propose an armonic expansion for the fields
of a form similar to (\ref{fluctarmonicsb}).
\beq
h (x,y) =\Sigma_I h^I (x) Y^I (y),\;\;
h_{(\alpha\beta)} =\Sigma_I b^I_{(\alpha\beta)}(x)Y^I(y),\;\;  
\delta\phi (x,y)= \Sigma_I f^I (x) Y^I (y)
\label{fluctarmonics2}
\eeq
Keeping only the s-wave fluctuations as before, 
we find after many computations 
(that are carefully spelled out in Appendix B), that like in IIB 
case, 
the many equations nicely combine and is sufficient to solve just 
one equation that reads,
\beq
\nabla^2 h^I + 6 g^{rr}\partial_r\phi 
\partial_r h^I + \frac{e^{3/2\phi}}{8}\Big( 31  F^2 +  
 57  (g^{\alpha\beta} F_{\alpha *}
F_\beta^*
- \frac{71}{57} g^{rr} F_{r *} F_r^*)\Big)h^I =0
\label{eq2a}
\eeq
As we have done in the type IIB section, the numerics in this case can 
be studied. We will not do this here and we just want to point to 
the fact 
that even when this has to be done in a completely numerical way (since 
the metric functions are only numerically known), there is a nice feature 
of this IIA solution. The fact that the dilaton does not diverge, 
makes us 
believe that the regularization will not be necessary. This is left for 
future work. The point of this section was just to call the reader's 
attention to this set of IIA models and show the analogy with the IIB 
treatment.

\section{Summary, Conclusions and Future Directions}

Let us start by summarizing
what we have done in this paper. First, we presented two 
Supergravity backgrounds (one in IIB and another in IIA) that are 
argued to be dual to $N=1$ SYM at low 
energies and we analyzed the spectrum in detail. Then, we initiated 
the study of glueball-like excitations in the strong coupling field 
theory as fluctuations of the Supergravity fields.

We presented the equations to study the spectrum of 
glueballs and their excitations and analyzed them
numerically. The type IIB case is analyzed in detail
and we proposed a regularization procedure that might be 
useful in other computations involving these wrapped branes set-ups.
Two appendixes present our computations in full detail.

We find that unlike some IIB backgrounds  previously  studied, 
in the D5 wrapped on $S^2$  model not even the simplest scalar mode 
decouples from the rest of the fluctuations. Indeed, as we have shown, 
assuming only fluctuations of the dilaton leads to inconsistent equations. 
Therefore,  the glueball  $0^{++}$ in the IIB model we studied, is  not dual to the dilaton, 
but to a mixture 
of dilaton and trace of the internal 
part of the metric. This goes in the same line as the papers 
\cite{Brower:1999nj}, where the glueballs turned out to be mixings 
between different Supergravity modes. 
This mixing might persist for higher spin modes. 
The presence of a non-constant dilaton background seems to be 
the reason for the mixing of the fluctuations and this also appears in 
the model studied in \cite{Gabadadze:2004jq}.  
Another important point is that the potential 
found produces a discrete spectrum and  
a mass gap without any sort of  cut-off, 
which seems to indicate that it is indeed capturing the physics of a 
confinig theory. As expected in a background with a linear dilaton, 
the states are not normalizable. 
Given that the UV completion of this theory is a little string theory 
it is not a  surprise that a regularization (or substraction) procedure is needed. 
We present a proposal for this regularization which amounts 
to substracting the unwanted contribution of the UV regime.

In the analysis of the type IIA background, we find that the scalar  mode does not decouple from 
the rest of the fluctuations, indicating that, 
indeed, the non-constant dilaton in the background
plays a role in producing this mixing. 
We do not perform the numerical analysis of the IIA background since the point of the 
paper is more to show a way of proceeding in these wrapped brane 
set-ups, that we believe is not exploited in the previous 
literature.

Let us now discuss some future directions to follow and some work 
that should be interesting to do in detail, not discussed in this 
paper.

First of all, as we mentioned in section 2, these models are 
contamined by the so called KK modes and we do not distinguish here 
if the glueballs we are obtaining are actually ``made out'' of KK 
modes composites (in which case they are not proper ``glueballs'' 
but hybrids composed of gluon, gluino and KK state)

A good technical way to distinguish is to repeat the computation we 
are doing in Section 3, but for the case of the ``dipole deformed'' 
field theory (see \cite{Gursoy:2005cn} for all details). Indeed, the 
idea in the paper \cite{Gursoy:2005cn} is to make a $SL(3,R)$ 
deformation of 
the supergravity solution, that reflects on the dual field theory 
side on particular deformation that affects only the dynamics of the 
KK modes in the spectrum. Hence, the comparison of the eqs 
(\ref{final}) and those in the appendix with those obtained in using 
the deformed metric can illuminate on what type of 
composition our `glueball' has, if only glue and gluino, or if it 
is composed out of KK modes. Same could be repeated with 
deformations of the $G_2$ holonomy manifolds and the many examples 
already existing in the literature. 

It would be nice to check how our regularizing procedure works with the 
``flavor branes'' introduced in \cite{Nunez:2003cf} when doing a dual to 
quenched SQCD. Indeed, there the 
idea was precisely the same, taking away from the computation the effects 
of the unwanted UV region. This can be understood by looking at the 
plots in  figures 2 and 3 of \cite{Nunez:2003cf}.

It should also be of interest, to study the glueballs in the type 
IIA model discussed in Section 4.1 of the paper \cite{Nunez:2001pt}.
This solution is basically the same as the one we discussed in the 
IIB section, after some dualities. So, the interest of studying 
glueballs in this case is obviously to see if the same spectrum is 
obtained, analyze differences among eqs of motion, etc.

Besides, it might have some interest to study the IIA case in more 
detail, not only numerically, as we pointed out above, but also  
from a  $G_2$ holonomy perspective. Our dilaton-metric-gauge field 
fluctuations must combine in some way in a pure metric fluctuation 
in eleven dimensions. It should be nice to see how this works.

Other models where it might have be interesting to apply our 
techniques (mainly the sort of manipulations explained in the 
Appendix B) is in models of non-supersymmetric duality. There is 
indeed one very clear model, that was studied in detail in the 
papers \cite{Bak:2003jk}.

One might also think about studying the `fermionic counterpart' of
what we have done in this paper, by fluctuating the fermionic fields
around the bosonic background.

On other respect, the comparison with the results from Lattice 
SYM should be done. There are some of these results in 
but we believe that the topic will evolve to allow better 
understanding and the contribution of this paper might be useful
in the comparison with lattice results. See for 
example\cite{Feo:2004mr}. Regarding this point, it should be interesting 
to study the spectrum of fermionic fluctuations (with fermionic fields 
vanishing in the background). This does not seem to have been exploited in 
the AdS/CFT literature, while other methods based on 
Veneziano-Yankielowicz and extensions, seem to give nice results 
\cite{Feo:2004mr}. This situation might clearly improve with some study.

The interest of studying glueballs goes beyond the simple fact of 
getting a discrete spectrum (that is by itself of enough interest). 
Indeed, glueballs play an important role in some recent advances regarding the study of Deep Inelastic and 
other types of Scattering 
using AdS/CFT techniques \cite{Polchinski:2002jw}. The knowledge of 
glueballs masses and 
profiles in different models might help to extend the results in papers 
like 
\cite{Polchinski:2002jw} to other `more realistic' models.
\section{Acknowledgments:} 
We thank Richard Brower, Jos\'e Edelstein, Nick Evans,  
Sean Hartnoll, Rafael Hern\'andez, Oliver Jahn, Martin Kruczenski, 
Juan Martin Maldacena, Alfonso Ramallo, Angel Paredes, 
Chung-I-Tan, Pere Talavera and Stathis Tompaidis
for discussions and comments that helped improving 
the presentation and interest of 
the results of this paper. Elena C\'aceres would 
like to thank the Theory Group at the 
University of Texas at Austin for hospitality 
during the final stages of this work.
This work was
supported  in part by the National Science Foundation under
Grant No. PHY-0071512 and PHY-0455649, the US Navy, Office of Naval Research, Grant Nos. N00014-03-1-0639 and N00014-04-1-0336, Quantum Optics
Initiative and  by funds
provided by the U.S.Department of Energy (DoE) under cooperative
research agreement DF-FC02-94ER408818. Elena C\'aceres is also 
supported by Mexico's Council of Science and Technology, CONACyT, 
grant No.44840.  Carlos Nu\~nez is a Pappalardo
Fellow.

\appendix
\section{Appendix: Some Geometrical identities}
\renewcommand{\theequation}{A.\arabic{equation}}
\setcounter{equation}{0}
In the following, we list some geometrical identities that were used 
in the derivation of the fluctuated eqs 
(\ref{deltarmunu3})-(\ref{delatmaxwell})
\bea
& &\delta R= \nabla_\mu\nabla_\nu h^{\mu\nu} -\nabla^2 h_\mu^\mu -
R_{\mu\nu}h^{\mu\nu},\nonumber\\
& &\delta 
R_{\mu\nu}=\frac{1}{2}[\nabla_\alpha\nabla_\mu h_{\nu}^{\alpha} +
\nabla_\alpha\nabla_\nu h_{\mu}^{\alpha} - \nabla_\nu\nabla_\mu 
h_{\alpha}^{\alpha}
- \nabla^2 h_{\mu\nu}  ], \;\;\sqrt{g+h} = \sqrt{g}(1 + 
\frac{\epsilon}{2} 
h_\mu^\mu ) + O(\epsilon^2),\nonumber\\
& & \delta \Gamma_{\mu\nu}^\lambda =
\frac{1}{2} g^{\alpha\lambda} (\nabla_\mu
h_{\alpha \nu} + \nabla_\nu
h_{\alpha \mu} - \nabla_\alpha
h_{\mu \nu} )\to 
 g^{\alpha\beta} \delta \Gamma_{\alpha\beta}^r = -\frac{1}{2} 
g^{rr} (\nabla_r h_{\mu}^\mu - 2 \nabla_\mu h_r^\mu)
\nonumber\\
& & \delta(\nabla^2\phi)= \nabla^2\delta\phi - g^{ab}\delta 
\Gamma_{ab}^\rho \partial_\rho \phi - h^{ab}\nabla_a \nabla_b \phi
\label{vargeom}
\eea

\section{Appendix: Derivation of the eqs in the IIB and IIA cases}
\renewcommand{\theequation}{B.\arabic{equation}}
\setcounter{equation}{0}
In this Appendix we fill in all the details that were left out in the 
computations 
that lead to eqs. (\ref{final}) and (\ref{eq2a}) in sections three and four.
\subsection{Glueballs with the D5 branes solution}
Let us start with the type IIB solution.
To study glueballs, we need the 
variation of the eqs of motion.
In the type IIB case, for the solution of D5 branes wrapping $S^2$, the Einstein 
eqs for the metric, dilaton and three form read, 
\beq
R_{\mu\nu}=\frac{1}{2}\partial_\mu\phi\partial_\nu\phi 
+\frac{g_s e^{\phi}}{4}\Big( F_{\mu**} F_\nu^{**} - 
\frac{1}{12}g_{\mu\nu}F_3^2   \Big)
\label{riccieq2ba}
\eeq
by contracting we get the Ricci scalar eq, 
\beq
R=\frac{1}{2}(\partial_\mu \phi)^2 + \frac{g_s e^\phi}{24}F_3^2 .
\label{ricciscalareq2ba}
\eeq
The dilaton, Maxwell and Bianchi eqs. are,
\beq
\nabla^2\phi= \frac{g_s e^\phi}{12}F_3^2,\;\;\partial_\mu\Big( 
\sqrt{g} e^\phi F^{\mu\nu\rho} \Big)=0,\;\; 
\partial_{[\mu} F_{\kappa \nu\rho]}=0
\label{dilmax2ba}
\eeq

Now, let us study the fluctuations of these eqs. above.
Let us assume that the background fields vary according to
\beq
g_{\mu\nu}\to g_{\mu\nu}+ \epsilon h_{\mu\nu},\;\; \phi\to 
\phi+\epsilon\delta\phi, \;\;F_{\mu\nu\rho}\to 
F_{\mu\nu\rho}+\epsilon \delta F_{\mu\nu\rho}
\label{fluct2ba}
\eeq
So, keeping only linear order in the parameter $\epsilon$, we get 
eqs for the fluctuated fields that read, for variation in 
the Ricci tensor eq. (\ref{riccieq2b}),
\bea
& &  \frac{1}{2}[\nabla_\alpha\nabla_\mu h_{\nu}^{\alpha} +
\nabla_\alpha\nabla_\nu h_{\mu}^{\alpha} - \nabla_\nu\nabla_\mu 
h_{\alpha}^{\alpha}
- \nabla^2 h_{\mu\nu}     ] = \frac{1}{2}[\partial_\mu \phi 
\partial_\nu \delta\phi +
\partial_\mu \delta\phi \partial_\nu \phi ]\nonumber\\
& & + \frac{g_s e^{\phi}}{4}
[\delta\phi F_{\mu **}F_{\nu}^{**} +\delta F_{\mu **} F_\nu^{**} +
F_{\mu **} \delta F_\nu^{**} - 2 h^{ab}F_{\mu a *}F_{\nu b }^* ] 
\nonumber\\
& & -\frac{g_s e^{\phi}}{48}[ g_{\mu\nu}( 2 F_3 \delta F_3 - 3 
h^{ab} 
F_{a**}F_b^{**} + \delta\phi
F_3^2) + h_{\mu\nu} F_3^2] 
\label{deltarmunu3} 
\eea
and for the Ricci scalar eq.(\ref{ricciscalareq2ba})
\bea
& &\nabla_\mu\nabla_\nu h^{\mu\nu} -\nabla^2 h_\mu^\mu -
R_{\mu\nu}h^{\mu\nu}= g^{\mu\nu}\partial_\mu\phi 
\partial_\nu\delta\phi -\frac{h^{\mu\nu}}{2}\partial_\mu\phi
\partial_\nu\phi+
\nonumber\\ & & \frac{g_s e^{\phi}}{24}[2 F_3 \delta F_3 -
3 h^{kl} F_{k **} F_l^{**} + \delta\phi F_3^2],
\label{riccivaried2b} 
\eea
For the fluctuated dilaton eq. we have,
\beq
\nabla^2\delta\phi - h^{\mu\nu}\nabla_\mu\partial_\nu\phi 
-\frac{1}{2} g^{rr}\partial_r\phi(2\nabla^\mu h_{r\mu} - 
\partial_rh_\mu^\mu) -\frac{g_s e^\phi}{12}(\delta\phi F^2 + 2 F 
\delta F - 3 h^{\mu\nu}F_{\mu **}F_\nu^{**})=0
\label{dileq}
\eeq
and for the fluctuation of the Maxwell eq. and Bianchi identity,
\eqn{delatmaxwell}{\partial_\mu[\sqrt{g}e^\phi ( (\frac{1}{2} 
h_\rho^\rho
+\delta\phi)F^{\mu\nu\alpha}+\delta F^{\mu\nu\alpha} - h^{\alpha 
c}F^{\mu\nu}_c + h^{\nu c}F^{\mu\alpha}_c -
h^{\mu c} F_c^{\nu\alpha})]=0, \;\;d\delta F=0}
Notice that the Ricci scalar eq. (\ref{riccivaried2b}) can be 
obtained by contracting the 
Ricci tensor eq. (\ref{deltarmunu3}) and substracting 
$h^{\mu\nu}R_{\mu\nu}$, so, in the 
following we will work with a contracted version of 
(\ref{deltarmunu3}).
In the derivation of these eqs, we have used the geometrical 
identities reviewed in Appendix A.

Now, let us assume a fluctuation for the metric of the form
\bea
& & h_{i j} = \frac{\lambda}{5} g_{i j} h^\alpha 
_\alpha,\;\;(i,j=\vec{x},t,r) \nonumber\\
& & h_{\alpha \beta } = h_{(\alpha \beta)} + \frac{1}{5} g_{\alpha \beta } 
h^{a}_{a},\;\;(\alpha\beta)=(\theta,\tilde{\theta},\varphi,
\tilde{\varphi}, \psi)
\label{shiftsa}
\eea
Where $\lambda$ is a constant and we  have performed a
shift in the $h_{\alpha\beta}$, this shift is proportional to the trace of
the internal part of the metric $h^\alpha_\alpha$.
Note that  the trace of the metric fluctuation over all the space 
is, $h^\mu_\mu = (\lambda +1)  h^\alpha _\alpha $.
Now, let us study the form of the fluctuation of the eq. that is 
obtained by contracting (\ref{deltarmunu3}) with the background 
metric $g^{\mu\nu} \delta R_{\mu\nu}
$,
\beq
\nabla_\mu\nabla_\nu h^{\mu\nu} -\nabla^2 h_\mu^\mu=
\partial^r \delta\phi \partial_r \phi  +\frac{g_s e^\phi}{24}
[(\delta\phi -\frac{1}{2}g^{\mu\nu}h_{\mu\nu})F_3^2  +
3 h^{\rho\sigma} F_{\rho**}F_\sigma^{**} +2\delta F_3\cdot F_3]\\
\eeq
and using (\ref{shiftsa}),
\bea
& & \frac{1}{5}\lambda\nabla^2_x  h^\alpha_\alpha
-(\lambda+1)(\nabla^2_x +\nabla^2_y) h^\alpha_\alpha- 
g^{rr}\partial_r \delta\phi \partial_r \phi  -
\frac{g_s e^\phi}{24}\Big[(\delta\phi 
-\frac{(\lambda+1)}{2}h^\alpha_\alpha)F_3^2\nonumber\\
& & +\frac{3}{5}(\lambda g^{rr}F_{r**}F_r^{**} 
+g^{ab}F_{a**}F_b^{**})h_\alpha^\alpha + 2 F_3 \delta F_3 
+3h^{(ab)}F_{a**}F_b^{**}\Big]=0.
\label{riccicont2b}
\eea
For the dilaton eq. (\ref{dileq}) we will have,
\bea
& & \nabla^2\delta\phi 
+(\frac{3\lambda+5}{10})g^{rr}\partial_r\phi\partial_r 
h_{\alpha}^\alpha - \frac{g_s e^\phi}{12}\Big[(\delta\phi 
+\frac{\lambda}{5}h_{\alpha}^\alpha)F^2 + 2 F_3 \delta 
F_3 -\nonumber\\
& &   -\frac{3}{5}h_{\alpha}^\alpha (\lambda g^{rr}F_{r**}F_r^{**} 
+ g^{ab}F_{a**}F_b^{**}) - 4 
h^{(\alpha\beta)}F_{\alpha**}F_\beta^{**}  \Big]
\label{eqdil2b}
\eea
We have used the eq. of motion for the dilaton in the background 
eq.(\ref{dilmax2b}) above. The 
next step is to introduce an expansion in harmonics for each of 
the fields
\beq
h^\alpha_\alpha (x,y) =\Sigma_I \Pi^I (x) Y^I (y),\;\;
h_{(\alpha\beta)} =\Sigma_I b^I(x)Y^I_{(\alpha\beta)}(y),\;\;  
\delta\phi (x,y)= \Sigma_I f^I (x) Y^I (y)
\label{fluctarmonicsa}
\eeq
In order to satisfy the Bianchi identity we write the fluctuation $\delta F_3$ in the form 
\beq
\delta F_{\mu\nu\rho}=\partial_{[\mu}\delta A_{\nu\rho]}
\eeq
 We will show that it is possible to find a  
fluctuation $\delta F_3$ orthogonal to the background $F_3$, {\it i.e}  $\delta F_3 \cdot F_3 =0 $, that satisfies Maxwell's equation.

Let us  write the different componentes of Maxwell's equation for a general fluctuation $\delta F_3=\partial_{[\mu}\delta A_{\nu\rho]}$ without demanding yet  orthogonality with the backgorund $F_3$. We get

$\bullet \nu,\alpha= a,b$ (angular)
\bea
&&\partial_r[\sqrt{g}e^\phi ( (\frac{1}{2} h_\mu^\mu 
+\delta\phi)F^{r a b }+ \partial^{[r}\delta A^{ ab]} - h^{b c}F^{ra}_c + h^{a c}F^{rb}_c - 
h^{r c} F_c^{ab})] \nonumber \\
&&+\partial_\theta[\sqrt{g}e^\phi ( (\frac{1}{2} h_\mu^\mu 
+\delta\phi)F^{\theta a b }+\partial^{[\theta}\delta A^{ ab]} - h^{b c}F^{\theta a}_c + h^{a c}F^{\theta b}_c - 
h^{\theta c} F_c^{ab})] +\partial_x[\sqrt{g}e^\phi ( \partial ^{[x} A^{ ab]} )] =0\nonumber\\.
\label{maxab}
\eea 

$\bullet \nu,\alpha= r,b$ (r,angular)
\bea
\partial_\theta[\sqrt{g}e^\phi ( (\frac{1}{2} h_\mu^\mu 
+\delta\phi)F^{\theta r b }+\partial ^{[\theta}\delta A^{ rb]} - h^{b c}F^{\theta r}_c + h^{r c}F^{\theta b}_c - 
h^{\theta c} F_c^{rb})] +\partial_x[\sqrt{g}e^\phi ( \partial ^{[x} \delta A^{ r b]} )]= 0 \nonumber\\.
\label{maxrb}
\eea

$\bullet \nu,\alpha= yz$ 
\bea
\partial_r[\sqrt{g}e^\phi ( \partial^{[r} \delta A^{yz]} )] 
+\partial_\theta[\sqrt{g}e^\phi (\partial^{[\theta}A^{ yz]} )] 
+\partial_x[\sqrt{g}e^\phi ( \partial^{[x}\delta A^{ yz]} )] =0.
\label{maxyz}
\eea 

$\bullet \nu,\alpha=(r,z)$
\bea
\partial_\theta[\sqrt{g}e^\phi ( \partial^{[\theta} A^{ rz]} )] 
+\partial_x[\sqrt{g}e^\phi ( \partial^{[x}\delta A^{ r z]} )] =0.
\label{maxrz}
\eea

Now demand  $\delta F_3 \cdot F_3 =0$, thus $\delta F_{rab}=\delta 
F_{\theta ab}=0 $. From equations (\ref{maxab}) and (\ref{maxrb}) above it is clear that a fluctuation of the form
\bea
-\partial_x[\sqrt{g}e^\phi ( \partial ^{[x} A^{ ab]} )]& =&\partial_r[\sqrt{g}e^\phi ( (\frac{1}{2} h_\mu^\mu 
+\delta\phi)F^{r a b } - h^{b c}F^{ra}_c + h^{a c}F^{rb}_c - 
h^{r c} F_c^{ab})] \nonumber \\
&+&\partial_\theta[\sqrt{g}e^\phi ( (\frac{1}{2} h_\mu^\mu 
+\delta\phi)F^{\theta a b } - h^{b c}F^{\theta a}_c + h^{a c}F^{\theta b}_c - 
h^{\theta c} F_c^{ab})]  \nonumber\\
-\partial_x[\sqrt{g}e^\phi ( \partial ^{[x} \delta A^{ r b]} )&=&\partial_\theta[\sqrt{g}e^\phi ( (\frac{1}{2} h_\mu^\mu 
+\delta\phi)F^{\theta r b } - h^{b c}F^{\theta r}_c + h^{r c}F^{\theta b}_c - 
h^{\theta c} F_c^{rb})] \nonumber\\
\eea 

with the other components  given by (\ref{maxyz}) and (\ref{maxrz}) will satisfy Maxwell's equation,
the Bianchi identity  and is such that  $\delta F_3 \cdot F_3 =0 $.

Therefore, we have only eqs. (\ref{riccicont2b}) and (\ref{eqdil2b}), that 
keeping only the S-wave in the 
expansion in harmonics (\ref{fluctarmonicsa}) take the form,
\bea
& & -(\frac{4\lambda}{5}+1)\nabla^2_x  \Pi(x)-
\partial^r f(x) \partial_r 
\Phi -\frac{g_s e^\phi}{24}\Big[(f(x) 
-\frac{(\lambda+1)}{2}\Pi(x))F_3^2\nonumber\\
& & +\frac{3}{5}(\lambda g^{rr}F_{r**}F_r^{**} 
+g^{ab}F_{a**}F_b^{**})\Pi(x) \Big] =0
\label{Pi}
\eea
and 
\bea
& & \nabla^2 f(x)+ (\frac{3\lambda}{10} +\frac{1}{2})\partial^r
\Pi(x) \partial_r \Phi -\frac{g_s}{12} e^\phi
\Big[ (f(x) + \frac{\lambda}{5}\Pi(x)) F_3^2 -\nonumber\\
& &\frac{3}{5}
(g^{ab}F_{a**}F_b^{**} + \lambda g^{rr}F_{r**}F_r^{**} 
)\Pi(x) \Big]=0 
\label{f}
\eea
Equations (\ref{Pi}) and (\ref{f}) look suggestively similar
so we will first check if there is any value of $\lambda$ for which 
they are the same.
Indeed, for the two equations to be equal we need to satisfy
\bea
& & 2(\frac{4\lambda}{5}+1)\Pi(x) = f(x),\;\;\nonumber\\
& & 2f(x) =(\frac{3\lambda}{10} +\frac{1}{2})\Pi(x), \nonumber\\
& & 
-(f(x) -\frac{(\lambda+1)}{2}\Pi(x))=f(x) 
+\frac{\lambda}{5}\Pi(x)
\label{coef3}
\eea
It is easy to check that $\lambda= -\frac{35}{29}$ does the job.
The eq. we need to solve is,

\eqn{finalap}{\nabla^2 f(x)+ 2g^{rr}\partial_r f(x)
\partial_r \phi +\frac{g_s}{12} e^\phi
\Big[ \frac{5}{2}F_3^2 +\frac{29\cdot 3}{10}
(g^{ab}F_{a**}F_b^{**} -\frac{35}{29}g^{rr}F_{r**}F_r^{**}) 
\Big]f(x) =0}

This is precisely the eq.(\ref{final}) we wanted to obtain. 
\subsection{Glueballs with the D6 branes solution}
The relevant background consists in $N$ D6 branes 
wrapping a three cycle inside a CY3 fold.
So,the metric of our  IIA solution in Einstein frame and the 
dilaton where given in 
(\ref{meteisnt2a}), and the Maxwell Field strength was,
\beq
F_2= k'(r) dr \wedge (d\hat{\psi} +\cos\theta d\varphi - 
\cos\tilde{\theta}d\tilde{\varphi}) - (k(r) +1) \sin\theta d\theta 
\wedge d\varphi + (k(r)-1) \sin\tilde{\theta} d\tilde{\theta}\wedge 
d\tilde{\varphi}, 
\eeq
with $$ k(r)= \frac{f^2 - c^2(1-g_3^2)}{f^2+ c^2(1+g_3)^2}.$$
Because of the 
many similarities 
with the IIB model studied in previous sections, we will use 
the same approach. Let us first study fluctuations.
The Lagrangian of IIA is
\bea
& & L=\sqrt{g}\Big(R-\frac{1}{2}(\partial\phi)^2 -\frac{1}{4} 
e^{3/2\phi} F_{2}^2 -\frac{1}{12}e^{-\phi}H_3^2 
-\frac{1}{48}e^{\phi/2}\hat{F}_4^2   \Big)\nonumber\\
& & +\frac{1}{2}B_2\wedge F_4\wedge F_4, \nonumber\\
& & \hat{F}_4= dC_3 - H_3\wedge A_1,\;\; F_2= dA_1, \;\; H_3= dB_2.
\label{2aa}
\eea
now, let us focus on the configurations that are of our interest, 
that is those where only the fields $\phi,\;g_{\mu\nu},\; A_\mu$ 
are turned on. The eqs of motion in this case are \footnote{Like in the main 
text of the paper, we take $g_s=1$ in this Appendix}
\bea
& & R_{\mu\nu}= \frac{1}{2}\partial_\mu\phi \partial_\nu\phi + 
\frac{1}{2}e^{3/2\phi}(F_{\mu *} F_\nu^* 
-\frac{1}{16}g_{\mu\nu}F^2),\nonumber\\
& & R= \frac{1}{2}(\partial\phi)^2 +\frac{3}{16}e^{3/2\phi}F^2,\;\; 
\nabla^2\phi = \frac{3}{8} e^{3/2\phi}F^2,\nonumber\\
& & \partial_\mu\Big(\sqrt{g} e^{3/2\phi}F^{\mu\nu}   
\Big)=0,\;\;\partial_{[\mu}F_{\nu\rho]}=0
\label{ecsmova}
\eea
Now, let us study the variations of these eqs. Under a fluctuation 
in all the relevant fields
\bea
& & \phi =\phi +\epsilon\delta\phi,\;\;\; g_{\mu\nu}=g_{\mu\nu} 
+\epsilon h_{\mu\nu}\;\;\; F_{\mu\nu}= F_{\mu\nu} + \epsilon \delta 
F_{\mu\nu}
\eea
we have to first order in the fluctuation parameter $\epsilon$,
\bea
& & \delta R_{\mu\nu}= \frac{1}{2}(\partial_\mu \phi 
\partial_\nu\delta\phi + \partial_\nu \phi
\partial_\mu\delta\phi ) +\frac{1}{2}e^{3/2\phi} \Big( 
\frac{3}{2}\delta\phi (F_{\mu *} F_\nu^* -\frac{1}{16} g_{\mu\nu} 
F^2) + F_{\mu *}\delta F_\nu^* + \nonumber\\
& & F_{\nu *}\delta F_\mu^* - 
h^{ab} 
F_{\mu a}F_{\nu b} -
\frac{1}{8} g_{\mu\nu} ( F \delta F 
-h^{\alpha\beta}F_{\alpha*}F_\beta^*  
) - \frac{1}{16}h_{\mu\nu}F^2 \Big), \nonumber\\
& & \delta R= g^{\mu\nu}\partial_\mu\phi \partial_\nu\delta\phi
-\frac{h^{\mu\nu}}{2}\partial_\mu\phi\partial_\nu\phi 
+\frac{3}{16}e^{3/2\phi}\Big( \frac{3}{2}\delta\phi F^2 + 2  F 
\delta F - 2 h^{\mu\nu}F_{\mu *}F_\nu^*  \Big), \nonumber\\
& & \delta(\nabla^2\phi)= \frac{3}{8}e^{3/2\phi} 
(\frac{3}{2}\delta\phi F^2 +  2  F
\delta F  -  2 h^{\mu\nu}F_{\mu *}F_\nu^*), \nonumber\\
& & 
\partial_\mu\Big(  \sqrt{g} e^{3/2\phi} [(3\delta\phi + 
h_{\mu}^{\mu}) \frac{F^{\mu\nu}}{2} +\delta F^{\mu\nu} - 
h^{\nu}_a F^{\mu a} -
h^\mu_a F^{\nu a}] \Big)=0
\label{varR2a}
\eea
So, using the geometrical variations given in the Appendix A
and  putting all together, we end up with three eqs, one for 
the variation of the dilaton, one for the Ricci scalar and the Maxwell 
eq. One can see that the eq for the variation of the Ricci tensor is 
included in the Ricci scalar variation.

So, we have
\beq
\nabla_\mu\nabla_\nu h^{\mu\nu} -\nabla^2 h_\mu^\mu= 
g^{\mu\nu}\partial_\mu\phi \partial_\nu\delta\phi
 +\frac{e^{3/2\phi}}{8}\Big( \frac{9}{4}\delta\phi F^2 + 3 F \delta 
F + h^{\mu\nu} F_{\mu *}F_\nu^* -\frac{h^\mu_\mu}{4} F^2  \Big)
\label{riccieq}
\eeq
\beq
\nabla^2\delta\phi - h^{\mu\nu}\nabla_\mu\nabla_\nu\phi 
-\frac{g^{rr}}{2}(2\nabla_\mu h_r^\mu -\nabla_r 
h_\mu^\mu)\partial_r\phi 
=\frac{3 e^{3/2\phi}}{8}\Big( \frac{3}{2}\delta\phi F^2 + 2 F \delta 
F - 2 h^{\mu\nu}F_{\mu *}F_\nu^*   \Big)
\label{dilatoneq}
\eeq
\beq
\partial_\mu\Big(  \sqrt{g} e^{3/2\phi} [(3\delta\phi + 
h_k^k) \frac{F^{\mu\nu}}{2} +\delta F^{\mu\nu} - h^{\nu}_a F^{\mu a} 
-h^{\mu}_a F^{a \nu}] \Big)=0
\label{maxwell}
\eeq
Now, let us propose a fluctuation of the metric $h_{\mu\nu}$ of the 
form
\beq
h_{ij}=\frac{\lambda}{5} h g_{ij}, (i,j=x,t,r) \;\;\; 
h_{\alpha\beta}= h_{(\alpha\beta)} +\frac{h}{5} 
g_{\alpha\beta} 
\label{metricflucta}
\eeq
Here, we denote the trace of the internal part of the metric as 
$h=h_a^a$. So, let us study the fluctuated eqs, we will have for the 
Ricci scalar,
\bea
& & \Big(\frac{(4\lambda +5)}{5}\nabla^2_x 
+\frac{(4 + 5\lambda)}{5}\nabla^2_y\Big)h 
+g^{rr}\partial_r\phi\partial_r\delta\phi +\frac{e^{3/2\phi}}{8} 
\Big( [\frac{9\delta\phi -(\lambda+1) h}{4}]F^2 + 3 F \delta F 
\nonumber\\
& &
+\frac{h}{5}(g^{\alpha\beta} F_{\alpha *} F_\beta^* 
+\lambda g^{rr} F_{r *} F_r^*)  \Big) 
-g^{(\alpha\beta)}(\nabla_\alpha\nabla_\beta h + F_{\alpha *} 
F_\beta^*)=0
\label{riccita}\eea
and for the dilaton, after the eq. of motion for the 
background dilaton field  has been used we will have, 
\bea
& & \nabla^2\delta\phi  
+(\frac{3\lambda + 5}{10}) g^{rr}\partial_r\phi \partial_r h 
- \frac{ e^{3/2\phi}}{8}\Big( [\frac{9}{2}\delta\phi 
+\frac{3\lambda}{5}h] F^2 + 6 F 
\delta F -\frac{6}{5}h (g^{\alpha\beta} F_{\alpha *} 
F_\beta^*
+\lambda g^{rr} F_{r *} F_r^*)\nonumber\\
& & - \frac{3}{4}g^{(\alpha\beta)}F_{\alpha 
*}F_\beta^*  \Big)=0
\label{dileq2}
\eea
and the Maxwell eq.
\bea
& & \partial_\mu\Big(  \sqrt{g} e^{3/2\phi} [(3\delta\phi + 
h) \frac{F^{\mu\nu}}{2} +\delta F^{\mu\nu} - 
\frac{\lambda}{5}g^{\nu r}F^\mu_r - g^{\nu a} h 
F^\mu_a - g^{(\nu a)} F^\mu_a +\nonumber\\
& & \frac{\lambda}{5}g^{\mu r} 
h F^\nu_r - g^{\mu a}h F^\nu_a +  
g^{(\mu a)} F^\nu_a] \Big)=0 
\label{maxwellfluct} 
\eea
Now, let us follow an analysis very similar to the one we have done 
for the Type IIB case. First, we will propose an expansion of the 
form (\ref{fluctarmonicsa}).
\beq
h (x,y) =\Sigma_I h^I (x) Y^I (y),\;\;
h_{(\alpha\beta)} =\Sigma_I b^I(x)Y^I_{(\alpha\beta)}(y),\;\;  
\delta\phi (x,y)= \Sigma_I f^I (x) Y^I (y)
\label{fluctarmonics2a}
\eeq
Then, let us impose that both eqs are 
indeed the same. For this to happen, we will also need that the 
fluctuation of 
the Maxwell field is orthogonal to the field itself, that is 
$$ \delta F_{\mu\nu}= \partial_{[\mu}\delta A_{\nu]}, \;\;\;F_2 
\delta F_2=0. $$ (the first part is to automatically solve the 
Bianchi identity) and we will also concentrate on the 
$s$-wave, that is all modes with higher harmonics in the spheres will 
not be 
considered, same for $g^{(\alpha\beta)}$ that contains higher 
harmonics. As before, the armonic decomposition is such that 
$\nabla^2_y h $ is composed of higher harmonics. Then 
dividing eq (\ref{dileq2}) 
by 6, we have that the following equalities have to be satisfied
\beq
f^I=6 (\frac{4\lambda +5}{5})h^I=(\frac{3\lambda+5}{60})h^I, \;\; 
3 f^I =(\frac{\lambda+1}{4} -\frac{\lambda}{10})h^I,
\eeq
for  both eqs will be equal. Indeed, we can see that they are 
solved by 
$$
\lambda= -\frac{71}{57}, \;\; 95 f =2 h,\;\;
$$
and that both eqs. (\ref{riccita}) and (\ref{dileq2}) actually read,
\beq
\nabla^2 h^I + 6 g^{rr}\partial_r\phi 
\partial_r h^I + \frac{e^{3/2\phi}}{8}\Big( 31  F^2 +  
 57  (g^{\alpha\beta} F_{\alpha *}
F_\beta^*
- \frac{71}{57} g^{rr} F_{r *} F_r^*)\Big)h^I =0
\label{eq2aa}
\eeq
Regarding the Maxwell eq, we can make an argument similar to the one 
in the IIB case .


\begin{thebibliography}{99}
\bibitem{Maldacena:1997re}
  J.~M.~Maldacena,
  Adv.\ Theor.\ Math.\ Phys.\  {\bf 2}, 231 (1998)
  [Int.\ J.\ Theor.\ Phys.\  {\bf 38}, 1113 (1999)]
  [arXiv:hep-th/9711200].


\bibitem{Gubser:1998bc}
  S.~S.~Gubser, I.~R.~Klebanov and A.~M.~Polyakov,
  Phys.\ Lett.\ B {\bf 428}, 105 (1998)
  [arXiv:hep-th/9802109].
  E.~Witten,
  Adv.\ Theor.\ Math.\ Phys.\  {\bf 2}, 253 (1998)
  [arXiv:hep-th/9802150].


\bibitem{Teper:1998kw}
  M.~J.~Teper,
  arXiv:hep-th/9812187.
  C.~J.~Morningstar and M.~J.~Peardon,
  Phys.\ Rev.\ D {\bf 60}, 034509 (1999)
  [arXiv:hep-lat/9901004].


\bibitem{Amsler:1995gf}
  C.~Amsler {\it et al.},
  Phys.\ Lett.\ B {\bf 342} (1995) 433.
  C.~A.~Meyer,
  AIP Conf.\ Proc.\  {\bf 698}, 554 (2004)
  [arXiv:hep-ex/0308010].


\bibitem{Kim:1985ez}
  H.~J.~Kim, L.~J.~Romans and P.~van Nieuwenhuizen,
  Phys.\ Rev.\ D {\bf 32}, 389 (1985).

\bibitem{Witten:1998zw}
  E.~Witten,
  Adv.\ Theor.\ Math.\ Phys.\  {\bf 2}, 505 (1998)
  [arXiv:hep-th/9803131].

\bibitem{Csaki:1998qr}
  C.~Csaki, H.~Ooguri, Y.~Oz and J.~Terning,
  JHEP {\bf 9901}, 017 (1999)
  [arXiv:hep-th/9806021].
  R.~de Mello Koch, A.~Jevicki, M.~Mihailescu and J.~P.~Nunes,
  Phys.\ Rev.\ D {\bf 58}, 105009 (1998)
  [arXiv:hep-th/9806125].
  M.~Zyskin,
  Phys.\ Lett.\ B {\bf 439}, 373 (1998)
  [arXiv:hep-th/9806128].
  H.~Ooguri, H.~Robins and J.~Tannenhauser,
  Phys.\ Lett.\ B {\bf 437}, 77 (1998)
  [arXiv:hep-th/9806171].
  J.~G.~Russo,
  Nucl.\ Phys.\ B {\bf 543}, 183 (1999)
  [arXiv:hep-th/9808117].
  C.~K.~Wen and H.~X.~Yang,
  arXiv:hep-th/0404152.
  K.~Suzuki,
  arXiv:hep-th/0411076.
  R.~C.~Brower, C.~I.~Tan and E.~Thompson,
  arXiv:hep-th/0503223.

\bibitem{Brower:1999nj}
  R.~C.~Brower, S.~D.~Mathur and C.~I.~Tan,
  Nucl.\ Phys.\ B {\bf 574}, 219 (2000)
  [arXiv:hep-th/9908196].
  R.~C.~Brower, S.~D.~Mathur and C.~I.~Tan,
  Nucl.\ Phys.\ Proc.\ Suppl.\  {\bf 83}, 923 (2000)
  [arXiv:hep-lat/9911030].
  R.~C.~Brower, S.~D.~Mathur and C.~I.~Tan,
  Nucl.\ Phys.\ B {\bf 587}, 249 (2000)
  [arXiv:hep-th/0003115].
  R.~C.~Brower, S.~D.~Mathur and C.~I.~Tan,
  arXiv:hep-ph/0003153.




\bibitem{Csaki:1998cb}
  C.~Csaki, Y.~Oz, J.~Russo and J.~Terning,
  Phys.\ Rev.\ D {\bf 59}, 065012 (1999)
  [arXiv:hep-th/9810186].
  J.~A.~Minahan,
  JHEP {\bf 9901}, 020 (1999)
  [arXiv:hep-th/9811156].
  J.~G.~Russo and K.~Sfetsos,
  Adv.\ Theor.\ Math.\ Phys.\  {\bf 3}, 131 (1999)
  [arXiv:hep-th/9901056].
  C.~Csaki, J.~Russo, K.~Sfetsos and J.~Terning,
  Phys.\ Rev.\ D {\bf 60}, 044001 (1999)
  [arXiv:hep-th/9902067].



\bibitem{Constable:1999ch}
  N.~R.~Constable and R.~C.~Myers,
  JHEP {\bf 9911}, 020 (1999)
  [arXiv:hep-th/9905081].
  N.~R.~Constable and R.~C.~Myers,
  JHEP {\bf 9910}, 037 (1999)
  [arXiv:hep-th/9908175].
  R.~Apreda, D.~E.~Crooks, N.~J.~Evans and M.~Petrini,
  JHEP {\bf 0405}, 065 (2004)
  [arXiv:hep-th/0308006].


\bibitem{Klebanov:2000hb}
  I.~R.~Klebanov and M.~J.~Strassler,
  JHEP {\bf 0008}, 052 (2000)
  [arXiv:hep-th/0007191].

\bibitem{Caceres:2000qe}
  E.~Caceres and R.~Hernandez,
  Phys.\ Lett.\ B {\bf 504}, 64 (2001)
  [arXiv:hep-th/0011204].
  M.~Krasnitz,
  arXiv:hep-th/0011179.
  L.~A.~Pando Zayas, J.~Sonnenschein and D.~Vaman,
  Nucl.\ Phys.\ B {\bf 682}, 3 (2004)
  [arXiv:hep-th/0311190].
  X.~Amador and E.~Caceres,
  JHEP {\bf 0411}, 022 (2004)
  [arXiv:hep-th/0402061].
  M.~Schvellinger,
  JHEP {\bf 0409}, 057 (2004)
  [arXiv:hep-th/0407152].
  E.~Caceres,
  arXiv:hep-ph/0410076.

\bibitem{Gubser:2004qj}
  S.~S.~Gubser, C.~P.~Herzog and I.~R.~Klebanov,
  JHEP {\bf 0409}, 036 (2004)
  [arXiv:hep-th/0405282].
  O.~Aharony,
  JHEP {\bf 0103}, 012 (2001)
  [arXiv:hep-th/0101013].

\bibitem{Ametller:2003dj}
  L.~Ametller, J.~M.~Pons and P.~Talavera,
  Nucl.\ Phys.\ B {\bf 674}, 231 (2003)
  [arXiv:hep-th/0305075].

\bibitem{Danielsson:1998wt}
  U.~H.~Danielsson, E.~Keski-Vakkuri and M.~Kruczenski,
  ``Vacua, propagators, and holographic probes in AdS/CFT,''
  JHEP {\bf 9901}, 002 (1999)
  [arXiv:hep-th/9812007].

\bibitem{Rey:1998ik}
  S.~J.~Rey and J.~T.~Yee,
  Eur.\ Phys.\ J.\ C {\bf 22}, 379 (2001)
  [arXiv:hep-th/9803001].


\bibitem{Maldacena:1998im}
  J.~M.~Maldacena,
  Phys.\ Rev.\ Lett.\  {\bf 80}, 4859 (1998)
  [arXiv:hep-th/9803002]. See also,
  S.~S.~Gubser, A.~A.~Tseytlin and M.~S.~Volkov,
  JHEP {\bf 0109}, 017 (2001)
  [arXiv:hep-th/0108205].
  N.~J.~Evans, M.~Petrini and A.~Zaffaroni,
  JHEP {\bf 0206}, 004 (2002)
  [arXiv:hep-th/0203203]. 

\bibitem{Maldacena:2000yy}
  J.~M.~Maldacena and C.~Nunez,
  Phys.\ Rev.\ Lett.\  {\bf 86}, 588 (2001)
  [arXiv:hep-th/0008001].
                                                                                
                                                                                
\bibitem{Chamseddine:1997nm}
  A.~H.~Chamseddine and M.~S.~Volkov,
  Phys.\ Rev.\ Lett.\  {\bf 79}, 3343 (1997)
  [arXiv:hep-th/9707176].
                                                                                
\bibitem{Nunez:2003cf}
  C.~Nunez, A.~Paredes and A.~V.~Ramallo,
  JHEP {\bf 0312}, 024 (2003)
  [arXiv:hep-th/0311201].


\bibitem{Gimon:2002nr}
  E.~G.~Gimon, L.~A.~Pando Zayas, J.~Sonnenschein and
M.~J.~Strassler,
  JHEP {\bf 0305}, 039 (2003)
  [arXiv:hep-th/0212061].
  R.~Apreda, F.~Bigazzi and A.~L.~Cotrone,
  JHEP {\bf 0312}, 042 (2003)
  [arXiv:hep-th/0307055].
  G.~Bertoldi, F.~Bigazzi, A.~L.~Cotrone, C.~Nunez and L.~A.~Pando
Zayas,
  Nucl.\ Phys.\ B {\bf 700}, 89 (2004)
F.~Bigazzi, A.~L.~Cotrone, L.~Martucci and L.~A.~Pando Zayas,
  Phys.\ Rev.\ D {\bf 71}, 066002 (2005)
  [arXiv:hep-th/0409205].
  F.~Bigazzi, A.~L.~Cotrone and L.~Martucci,
  Nucl.\ Phys.\ B {\bf 694}, 3 (2004)
  [arXiv:hep-th/0403261].
\bibitem{Andrews:2005cv}
  R.~P.~Andrews and N.~Dorey,
  arXiv:hep-th/0505107.

\bibitem{Bertolini:2003iv}
  M.~Bertolini,
  Int.\ J.\ Mod.\ Phys.\ A {\bf 18}, 5647 (2003)
  [arXiv:hep-th/0303160].
  F.~Bigazzi, A.~L.~Cotrone, M.~Petrini and A.~Zaffaroni,
  Riv.\ Nuovo Cim.\  {\bf 25N12}, 1 (2002)
  [arXiv:hep-th/0303191].
  E.~Imeroni,
  arXiv:hep-th/0312070.
  A.~Paredes,
  arXiv:hep-th/0407013.



\bibitem{Evans:2005ip}
  N.~Evans, J.~P.~Shock and T.~Waterson,
  arXiv:hep-th/0505250.







\bibitem{Gopakumar:1998ki}
R.~Gopakumar and C.~Vafa,
``On the gauge theory/geometry correspondence,''
Adv.\ Theor.\ Math.\ Phys.\  {\bf 3} (1999) 1415
[arXiv:hep-th/9811131].
                                                                                
\bibitem{Vafa:2000wi}
C.~Vafa,
``Superstrings and topological strings at large N,''
J.\ Math.\ Phys.\  {\bf 42} (2001) 2798
[arXiv:hep-th/0008142].
\bibitem{Seiberg:1997ax}
N.~Seiberg,
``Notes on theories with 16 supercharges,''
Nucl.\ Phys.\ Proc.\ Suppl.\  {\bf 67} (1998) 158
[arXiv:hep-th/9705117].
\bibitem{Edelstein:2001pu}
  J.~D.~Edelstein and C.~Nunez,
  JHEP {\bf 0104}, 028 (2001)
  [arXiv:hep-th/0103167].
\bibitem{Itzhaki:1998dd}
N.~Itzhaki, J.~M.~Maldacena, J.~Sonnenschein and S.~Yankielowicz,
``Supergravity and the large N limit of theories with sixteen
supercharges,''
Phys.\ Rev.\ D {\bf 58} (1998) 046004
[arXiv:hep-th/9802042].
                                                                                                                             
\bibitem{Atiyah:2000zz}
M.~Atiyah, J.~M.~Maldacena and C.~Vafa,
``An M-theory flop as a large N duality,''
J.\ Math.\ Phys.\  {\bf 42} (2001) 3209
[arXiv:hep-th/0011256].
M.~Atiyah and E.~Witten,
``M-theory dynamics on a manifold of G(2) holonomy,''
arXiv:hep-th/0107177.
\bibitem{Acharya:1998pm}
  B.~S.~Acharya,
  Adv.\ Theor.\ Math.\ Phys.\  {\bf 3}, 227 (1999)
  [arXiv:hep-th/9812205].
\bibitem{Acharya:2000gb}
B.~S.~Acharya,
``On realising N = 1 super Yang-Mills in M theory,''
arXiv:hep-th/0011089.
J.~Gomis,
``D-branes, holonomy and M-theory,''
Nucl.\ Phys.\ B {\bf 606} (2001) 3
[arXiv:hep-th/0103115].


\bibitem{Acharya:2001hq}
B.~S.~Acharya,
``Confining strings from G(2)-holonomy spacetimes,''
arXiv:hep-th/0101206.

\bibitem{Acharya:2001dz}
B.~S.~Acharya and C.~Vafa,
arXiv:hep-th/0103011.

\bibitem{Hartnoll:2002th}
  S.~A.~Hartnoll and C.~Nunez,
  JHEP {\bf 0302}, 049 (2003)
  [arXiv:hep-th/0210218].


\bibitem{Gursoy:2003hf}
  U.~Gursoy, S.~A.~Hartnoll and R.~Portugues,
  Phys.\ Rev.\ D {\bf 69}, 086003 (2004)
  [arXiv:hep-th/0311088].




\bibitem{Hartnoll:2004yr}
  S.~A.~Hartnoll and R.~Portugues,
  Phys.\ Rev.\ D {\bf 70}, 066007 (2004)
  [arXiv:hep-th/0405214].





\bibitem{Cvetic:2001ih}
  M.~Cvetic, G.~W.~Gibbons, H.~Lu and C.~N.~Pope,
  Phys.\ Rev.\ Lett.\  {\bf 88}, 121602 (2002)
  [arXiv:hep-th/0112098].



\bibitem{Brandhuber:2001kq}
  A.~Brandhuber,
  Nucl.\ Phys.\ B {\bf 629}, 393 (2002)
  [arXiv:hep-th/0112113].
\bibitem{Arean:2005ar}
  D.~Arean, A.~Paredes and A.~V.~Ramallo,
  arXiv:hep-th/0505181.

\bibitem{Gibbons:2002pq}
  G.~Gibbons and S.~A.~Hartnoll,
  Phys.\ Rev.\ D {\bf 66}, 064024 (2002)
  [arXiv:hep-th/0206202].


\bibitem{Gabadadze:2004jq}
  G.~Gabadadze and A.~Iglesias,
  Phys.\ Lett.\ B {\bf 609}, 167 (2005)
  [arXiv:hep-th/0411278].
A.~Hashimoto and Y.~Oz,
  Nucl.\ Phys.\ B {\bf 548}, 167 (1999)
  [arXiv:hep-th/9809106].

\bibitem{Gursoy:2005cn}
  U.~Gursoy and C.~Nunez,
  arXiv:hep-th/0505100. See also 
  S.~Pal,
  arXiv:hep-th/0505257.



\bibitem{Nunez:2001pt}
  C.~Nunez, I.~Y.~Park, M.~Schvellinger and T.~A.~Tran,
  JHEP {\bf 0104}, 025 (2001)
  [arXiv:hep-th/0103080].

\bibitem{Bak:2003jk}
  D.~Bak, M.~Gutperle and S.~Hirano,
  JHEP {\bf 0305}, 072 (2003)
  [arXiv:hep-th/0304129].
  D.~Z.~Freedman, C.~Nunez, M.~Schnabl and K.~Skenderis,
  Phys.\ Rev.\ D {\bf 69}, 104027 (2004)
  [arXiv:hep-th/0312055].
  A.~B.~Clark, D.~Z.~Freedman, A.~Karch and M.~Schnabl,
  Phys.\ Rev.\ D {\bf 71}, 066003 (2005)
  [arXiv:hep-th/0407073].
I.~Papadimitriou and K.~Skenderis,
  JHEP {\bf 0410}, 075 (2004)
  [arXiv:hep-th/0407071].
  D.~Bak and H.~U.~Yee,
  Phys.\ Rev.\ D {\bf 71}, 046003 (2005)
  [arXiv:hep-th/0412170].
  D.~Bak, M.~Gutperle, S.~Hirano and N.~Ohta,
  Phys.\ Rev.\ D {\bf 70}, 086004 (2004)
  [arXiv:hep-th/0403249].


\bibitem{Feo:2004mr}
  A.~Feo, P.~Merlatti and F.~Sannino,
  Phys.\ Rev.\ D {\bf 70}, 096004 (2004)
  [arXiv:hep-th/0408214].
  A.~Feo,
  Mod.\ Phys.\ Lett.\ A {\bf 19}, 2387 (2004)
  [arXiv:hep-lat/0410012].



\bibitem{Polchinski:2002jw}
  J.~Polchinski and M.~J.~Strassler,
  JHEP {\bf 0305}, 012 (2003)
  [arXiv:hep-th/0209211].
  J.~Polchinski and M.~J.~Strassler,
  Phys.\ Rev.\ Lett.\  {\bf 88}, 031601 (2002)
  [arXiv:hep-th/0109174].
  H.~Boschi-Filho and N.~R.~F.~Braga,
  Phys.\ Lett.\ B {\bf 560}, 232 (2003)
  [arXiv:hep-th/0207071].
  R.~C.~Brower and C.~I.~Tan,
  Nucl.\ Phys.\ B {\bf 662}, 393 (2003)
  [arXiv:hep-th/0207144].
  S.~J.~Brodsky and G.~F.~de Teramond,
  Phys.\ Lett.\ B {\bf 582}, 211 (2004)
  [arXiv:hep-th/0310227].
  O.~Andreev,
  Phys.\ Rev.\ D {\bf 70}, 027901 (2004)
  [arXiv:hep-th/0402017].
  H.~Nastase,
  arXiv:hep-th/0410124.
  K.~Kang and H.~Nastase,
  arXiv:hep-th/0501038.








\end{thebibliography}
\end{document}